\documentclass[times,sort&compress,3p]{elsarticle}
\journal{Journal of Multivariate Analysis}

\graphicspath{{./art/}}

\usepackage[plain,noend]{algorithm2e}
\usepackage{amsmath, amssymb, enumerate, amsthm,color}
\usepackage{algorithmic,extarrows,cleveref}
\usepackage{subcaption}

\crefformat{section}{\S#2#1#3} 
\crefformat{subsection}{\S#2#1#3}
\crefformat{subsubsection}{\S#2#1#3}

\newcommand {\dps}  {\displaystyle}
\newcommand {\Reals}  {{\rm I \! R}}
\newcommand {\Naturals}  {{\rm I \! N}}
\newcommand {\reals}  {{\rm I \! R}}

\newcommand{\1}{{\bf 1}}

\newcommand{\A}{\mathbf{A}}

\newcommand{\D}{\mathbf{D}}
\newcommand{\E}{\mathbf{E}}

\newcommand{\I}{\mathbf{I}}

\newcommand{\R}{\mathbf{R}}

\newcommand{\U}{\mathbf{U}}

\newcommand{\V}{\mathbf{V}}
\newcommand{\W}{{\mathbf{W}}}
\newcommand{\X}{{\mathbf{X}}}
\newcommand{\Y}{{\mathbf{Y}}}
\newcommand{\Z}{\mathbf{Z}}

\newcommand{\expc}{{\E}}

\newcommand{\cB}{\mathcal{B}}
\newcommand{\cC}{\mathcal{C}}
\newcommand{\cD}{\mathcal{D}}
\newcommand{\cF}{\mathcal{F}}

\newcommand{\cL}{\mathcal{L}}
\newcommand{\cS}{\mathcal{S}}

\newcommand{\cU}{\mathcal{U}}

\newcommand{\0}{{\vec{\mathbf{0}}}}
\renewcommand{\a}{{\mathbf{a}}}

\newcommand{\e}{{\mathbf{e}}}

\renewcommand{\u}{{\mathbf{u}}}

\newcommand{\x}{{\mathbf{x}}}
\newcommand{\y}{{\mathbf{y}}}

\newcommand{\cN}{{\cal N}}

\newtheorem{thm}{Theorem}
\newtheorem{lem}[thm]{Lemma}
\newtheorem{cor}[thm]{Corollary}

\newtheorem{assum}{Assumption}

\newcommand{\softmax}{{\text{softmax}}}

\newcommand{\diag}{{\rm diag}}
\newcommand{\Prob}{{\mathbf P}}

\newcommand{\DiaEig}{{ \Sigma }}

\newcommand{\ignore}[1]{}

\usepackage{lineno}

\journal{Journal of Multivariate Analysis}

\begin{document}
	\begin{frontmatter}
\title{Multi-sample estimation of centered log-ratio matrix in microbiome studies}

\author{Yezheng Li\corref{cor1}}
\ead{yezheng@alumni.upenn.edu}
\author{Hongzhe Li}
\author{Yuanpei Cao}

\address{Department of Biostatistics and Epidemiology, Perelman School of Medicine, University of Pennsylvania, Philadelphia, PA, United States}
\cortext[cor1]{Corresponding author}

	\begin{abstract}
		In microbiome studies, one of the ways of studying bacterial abundances is to estimate bacterial composition based on the sequencing  read counts. Various transformations are then applied to such compositional data  for downstream statistical analysis, among which the centered log-ratio (\textsc{clr}) transformation is most commonly used. 
		Due to limited sequencing depth and DNA dropouts, many rare bacterial taxa  might not be captured  in the final  sequencing reads, which results in many zero counts. Naive composition estimation using count normalization  leads  to many zero proportions, which makes  \textsc{clr} transformation infeasible. This paper proposes a multi-sample approach to estimation of the \textsc{clr} matrix directly in order to borrow information across samples and across species.  Empirical results from real datasets suggest  that the \textsc{clr} matrix over  multiple samples is approximately low rank, which  motivates a regularized maximum likelihood estimation with a nuclear norm penalty. An efficient optimization algorithm using the generalized accelerated proximal gradient is developed. Theoretical upper bounds  of  the estimation errors  and of its corresponding singular subspace errors are established. Simulation studies demonstrate that the proposed  estimator outperforms the naive estimators. The method is analyzed on \textcolor{black}{Gut Microbiome dataset} and \textcolor{black}{ the American Gut project}
		.
		  
	\end{abstract}
	

\end{frontmatter}

\section{Introduction}

Recent studies have demonstrated that the microbiome composition varies across individuals due to different health and environmental conditions \citep{creasy2012framework,cao2020multisample}. Microbiome is associated with many complex diseases such as obesity, atherosclerosis, and Crohn's disease \citep{turnbaugh2009core,koeth2013intestinal,lewis2015inflammation}. With  the development of next-generation sequencing technologies, the human microbiome can be quantified by using direct DNA sequencing of either marker genes or the whole metagenomes. After aligning the sequence reads to the reference microbial genomes, one obtains counts of sequencing reads that can be assigned to a set of bacterial taxa observed in the samples.  Such count data provide information about the relative abundance of different bacteria in different samples. 

Due to limited sequencing depths and DNA dropouts during sequencing, count results many zeros and therefore the relative proportional of bacterial taxa often include many zeros.  Excessive zeros in the proportions complicate many downstream data analyses. 
Since the pioneering work of \citep{aitchison1982statistical,aitchison1983principal,egozcue2003isometric}, several techniques have been proposed to deal with zeros in compositional or count data (see \citep{martin2011dealing} for an overview). When the data are compositional, they need to be scaled so that subsequent analysis are scale-invariant, and geometrically this means to force them into the open simplex. A common practice to analyze compositional data is to map bijectively the  compositions into the ordinary Euclidean space through a suitable transformation, so that standard multivariate analysis techniques can be used  \citep{aitchison1983principal,egozcue2003isometric}. Among many such transformations \citep{aitchison1983principal,egozcue2003isometric,andrews2015generalized}, the  center log-ratio (\textsc{clr}) tranformation, defined as the logarithms of the bacterial composition subtracted by logarithm of the geometric mean, is most widely used in practical analysis of microbiome data.   After such transformation,
one can then apply the standard statistical analysis methods such as the principal component analysis based on the \textsc{clr} transformed data \citep{aitchison1983principal,filzmoser2009principal}.

Since the original data observed are counts instead of compositions in microbiome studies, one has to first estimate the compositions  before applying the \textsc{clr} transformation. The most commonly applied methods in composition data analysis involve a two-step procedure. One  first estimates the composition using the observed count data and then performs the \textsc{clr} transformation \citep{martin2015bayesian,cao2020multisample}.  Since the counts often includes many zeros, such zeros can just be replaced by an arbitrarily small numbers so that one can furtherly apply  the \textsc{clr} transformation.  \ignore{\cite{cao2020multisample} developed a multi-sample based estimation of the compositional matrix by borrowing information across the samples. }
One drawback of estimating the \textsc{clr} matrix from the estimated compositions is that the uncertainty in the estimated compositions is not accounted when they are transformed  using the \textsc{clr}s. 

In this paper, we propose a method to estimate the \textsc{clr} matrix directly based on the observed count data.  One key idea of the proposed method is to estimate the \textsc{clr} matrix of compositions of mutiple samples together, i.e., the \textsc{clr} matrix estimated from the count  data from multiple samples. This effectively borrows information across multiple samples in order to obtain better estimate of the \textsc{clr} for each of the samples. 
More specifically, our proposed approach is based on a penalized likelihood estimation parameterized directly based on the \textsc{clr} matrix, where a nuclear norm penalty on the \textsc{clr} matrix  is imposed to capture the expected approximate low-rank structure of the \textsc{clr} matrix.  The low rank assumption is based on the empirical observations that the bacteria taxa abundances tend to be highly correlated and individual gut microbiome samples tend to cluster together to form discrete microbial communities.  This is different from the approach of \cite{cao2020multisample}, where the low-rank assumption is directly imposed on the compositional matrix. 
Since there is no constraints on the \textsc{clr} matrix (except trivial constraints that sum of each rows to be zero), we develop a   generalized accelerated proximal gradient algorithm to efficiently perform the optimization. The computation is faster than  that of  \cite{cao2020multisample} where a simplex projection step is needed to account for the  bounded simplex constraints. 

We obtain the estimation bounds of the proposed estimator and its corresponding singular vector under both the exact low-rank and approximate low-rank settings.   We present simulation results to compare our estimate and commonly used zero-replacement estimate. Finally, we demonstrate the methods using  the data set from  \cite{Wu2011} and data set from the  American Gut Project \citep{McDonalde00031-18}.

\section{A Poisson-Multinomial Model for Microbiome Count Data}
\ignore{For any integer $n>0$, we write $[n]=\{1,\ldots,n\}$ and denote $\e_i(n)$ as the canonical basis in $\reals^n$ with $i$th entry being  one and others being zero. }
We refer to any $u\in \mathbb{R}^p$ as a composition vector if $u_i > 0$ for  $i=1,2,\ldots, p$ and $\displaystyle\sum_{i=1}^p u_i = 1$.
The data observed in typical marker gene-based microbiome studies (i.e., 16S rRNA marker gene) can be summarized  as follows. Let  $N_i$ be the total number of sequencing reads for the $i$th sample that can be assigned to one of the $p$ bacterial taxa, and $W_{ij}$ be the read count that can be assigned to the $j$th taxon for $j=1,\cdots,p$,  where $\displaystyle N_i=\sum_{j=1}^p W_{ij}$.  It is natural to model the  count data $\W_i=(W_{ij}, j=1,\cdots, p)$  using  a multinomial distribution with composition parameter $\X_i^*=(X_{ij}, j=1,\cdots,p)$ with $\displaystyle \sum_{j=1}^p X_{ij}=1$ \citep{cao2020multisample}.  Let $\X^*=(\X^*_{ij}) \in \mathbb{R}^{n\times p}$ denote the $n\times p$ compositional matrix. 

Since each row of the compositional matrix $\X$ ($\X$ can be true parameter $\X^*$ or estimated one $\hat {\X}$) is within the $p-1$ dimensional simplex with a unit sum constraint,  certain transformation is often needed for downstream statistical analysis, including principal component analysis, estimation of covariance and regression analysis. One of the transformations that has been widely used in compositional data analysis is the \textsc{clr} transformation \citep{aitchison1983principal,aitchison1982statistical}, which 
is defined   as $\Z_{ij}=\log (X_{ij}/g(\X_i))$ where $g(\a)=\left(\prod_{i=1}^pa_{i}\right)^{1/p}$ is the geometric mean of the $p$ proportions. This can be written as a vector form as 
$$
\Z_i=\textsc{clr}(\X_i)=\log \X_{i}  \cdot  \left(\I_p - \frac1p \1_p\1_p\right).
$$
The inverse of the \textsc{clr} transformation, which returns the original compositional vector $\X^*_i$, is  actually the softmax function defined as 
$$
\X_i= \textsc{clr}^{-1}(\Z_i) = \softmax(\Z_i) = \left(\frac{\exp\left(Z_{ij}\right)}{ \sum_{k =1 }^{p }\exp\left(Z_{ik}\right)}\right)_{n \times p},
$$
and the gradient of the softmax function is $$\displaystyle\nabla \softmax (\Z_i) = \diag\left\{\softmax (\Z_i) \right\} - \left[\softmax(\Z_i) \right]^{T} \softmax(\Z_i	)
\in \Reals^{p \times p}.$$
We let $\Z^*=(\Z^*_{ij})\in \mathbb{R}^{n\times p}$ denote the  matrix of the underlying true centered log-ratio transformation of $n$ samples  over $p$ taxa.  Different from the work focusing on estimating $\X^*\in \Reals^{n \times p}$ \citep{cao2020multisample,martin2015bayesian}, our goal is to estimate this \textsc{clr} matrix $\Z^*$ based on the observed counts $\W\in \Reals^{n \times p }$.

Using the \textsc{clr} matrix $\Z^*$ as the parameter, the proposed Poisson-multinomial model for count-compositional data can be written as 
\begin{eqnarray}
N_i &\sim &\text{Pois}(\nu_i), i =1,2,\ldots, n; \nonumber \\
 	\Prob_{\Z_i^*}\left( W_{i1}, \ldots, W_{ip}| N_i \right) &= &\frac{N_i!}{\prod_{j=1}^p W_{ij}!} \prod_{j = 1}^p \left( X^*_{ij}\right)^{W_{ij}} , i =1,2,\ldots, n. 
\label{eq:multinomial distribution}
\end{eqnarray}
where $X_{ij}^* =\softmax \left( \Z_i^* \right)_j = \textsc{clr}^{-1}\left( \Z_i^* \right)_j$.

The maximum likelihood estimation (of each composition vector in each row) provides one naive estimation $\hat{\Z}^{\rm MLE}$ of the \textsc{clr} matrix $\Z^*$, which is equivalent to estimating each row $\Z^*_i \in \Reals^{1\times p}$ separately using only the data observed for  the $i$th sample.  However, $\hat{\Z}^{\rm  MLE}$ cannot resolve zero-count issue: $\hat{Z}_{ij}^{\rm  MLE} = -\infty$ and then $\softmax\left(\hat{\Z}^{\rm  MLE} _i\right)_{j}= 0$ when $W_{ij}= 0 $. One standard and commonly used method of avoiding assigning zeros to $\hat{X}_{ij}$ is  the zero-replacement estimation $\hat{\X}^{zr}$:
$$
\hat{X}_{ij}^{\rm zr}= \frac{W_{ij} \wedge a}{\sum_{j=1}^{p} \left(W_{ij}\wedge a\right) },
$$
where $a $ is an arbitrarily small number, but commonly set $a=0.5$ \citep{cai2016structured,cao2020multisample,aitchison1983principal,martin2011dealing,martin2003dealing}.

 On the other hand, empirical observations in real microbiome data suggest that the \textsc{clr} matrix $\Z^*$ or composition matrix $\X^*$ is usually approximate low-rank due to dependency among the bacterial taxa.  In this paper, we explore this low-rank structure $\Z^*$ to provide an improved estimate of $\Z^*$. This is different from   \cite{cao2020multisample}, where  composition matrix $\X^*$  is assumed to be approximate low-rank. 
\section{Regularized Estimation of the Centered-Log-Ratio Matrix and the Computational Algorithm}
\label{sec:regularized estimation}
\subsection{Regularized estimation of the centered-log-ratio matrix}
\label{subsec:regularized estimation with fixed tuning parameter}
In order to improve  the estimate of the \textsc{clr} matrix $\Z^*$,  the approximate low-rank structure of the $\Z^*$ is explored.  The co-occurrence patterns \citep{faust2012microbial}, various symbiotic relationships in microbial communities \citep{woyke2006symbiosis,horner2007comparison,chaffron2010global} and samples in similar microbial communities  are expected to lead to an approximately low-rank structure of the \textsc{clr} matrix in the sense that the singular values of $\Z^*$ decay to zero in a fast rate. Such a low-rank structure of $\Z^*$ is further investigated in our real data analysis in \cref{sec:real data analysis},  showing  the empirical evidence of approximate low-rank \textsc{clr} matrix. 
We propose  the following nuclear-norm penalized estimation of the \textsc{\textsc{clr}} matrix $\Z^*$ by exploring the low-rank structure of such a matrix, 
\begin{equation}
\hat \Z (\lambda)  \in \arg \min_{\Z \1_p = \0_n}  \cL_{N} \left(  \Z ;\W \right) + \lambda \left\| \Z\right\|_{*},\label{eq:nuclear norm estimation clr}
\end{equation}
where
{\allowdisplaybreaks
	\begin{eqnarray}
	\nonumber \cL_{N} \left(  \Z;\W  \right) & \triangleq &   - \frac{1}{N} \sum_{i=1}^{n} N_i \cL_{N_i}(\Z_i;\W)\triangleq   - \frac{1}{N} \sum_{i=1}^{n} \sum_{j=1}^{p } W_{ij} \log \left\{\softmax \left( \Z_i\right)_j \right\} \\
	\nonumber &=&- \frac{1}{N} \sum_{i=1}^{ n} \sum_{j=1}^{p } W_{ij} \log \left( \frac{ e^{ z_{ij}  }}{ \sum_{j=1}^{ p} e^{ z_{ij} } }\right) =\frac{1}{N} \sum_{i=1}^{n}\left\{N_i  \log \left(  \sum_{j=1}^{p } e^{ z_{ij}  } \right)  -  \sum_{j=1}^{p } W_{ij} z_{ij} \right\}. \label{eq:defn likelihood ratio}
	\end{eqnarray}
}

The proposed estimator \eqref{eq:nuclear norm estimation clr} is  a regularized nuclear norm minimization which can be solved by either semidefinite programming via interior-point semidefinite programming (SDP) solver, or first-order method via Templates for First-Order Conic Solvers (TFOCS). However the interior-point SDP solver computes the nuclear norm via a less efficient eigenvalue decomposition, which does not scale well with large $n$ and $p$. TFOCS on the other hand often results in the oscillations or overshoots along the trajectory of the iterations.

To achieve a stable and efficient optimization for \eqref{eq:nuclear norm estimation clr} with large $n$ and $p$, we propose an algorithm based on the generalized accelerated proximal gradient method and Nesterov's scheme. Compared to \cite{cao2020multisample} which focus on estimating $\X^*$ and introducing nuclear norm regularization of $\X^*$, we do not need further projections and the zero-sum  constraints of each row is  automatically  satisfied in our optimization algorithm. Algorithm with fixed tuning parameter $\lambda$ is in \cref{subsec:A generalized accelerated proximal gradient algorithm} and auto-tuning procedure is in \cref{subsec:selection of tuning parameters}. More details of  \cref{subsec:A generalized accelerated proximal gradient algorithm} and \cref{subsec:selection of tuning parameters} are provided in \cref{app:details of algorithms}.

\subsection{A generalized accelerated proximal gradient algorithm}
\label{subsec:A generalized accelerated proximal gradient algorithm}
We present an optimization algorithm for \eqref{eq:nuclear norm estimation clr} based on the generalized accelerated Nesterov's scheme, which follows the formulation of \cite{beck2009fast,cao2020multisample} and the spirit of \cite{su2014differential}. 

The algorithm involves the following steps: 
 First,  based on the count matrix, we initialize $\Z^{(0)}, \Y^{(0)} \in \Reals^{n \times p}$ as 
	\begin{eqnarray} \hat Z_{ij}^{(0)} & \leftarrow  & \hat {Z}_{ij}^{\rm zr} +\epsilon_{ij}  = \textsc{clr}\left( \frac{W_{ij} \wedge 0.5}{\sum_{j=1}^{p} \left(W_{ij}\wedge 0.5\right) } \right) +\epsilon_{ij} ,  \label{eq:zero replacement} \\ 
	\hat  Y_{ij}^{(0)}&  \leftarrow & \hat {Z}_{ij}^{(0)}. \nonumber
	\end{eqnarray}
	where $\epsilon \in \Reals^{n \times p}$ is the perturbation and $\epsilon =  \tilde{\epsilon }  \cdot  \left(\I_p - \frac1p \1_p\1_p\right)$ with $\tilde{\epsilon } \in \Reals^{n \times p}$  and summation of each row of $\epsilon $ is guaranteed to be zeros while $\tilde{\epsilon }$ have $n\times p $ independent and randomly-generated entries. 
It is worth noticing the perturbation $\epsilon$ does not appear in \cite{cao2020multisample} and theoretically is not needed  in convex optimization, but more likely to appear  in  non-convex optimization scenarios (for example, neural network scenarios).  However due to numerical instability of centroid-log-ratio and $\softmax$ function \citep{galletti2016numerical}, perturbation $\epsilon$ is important to ensure the  stability of the proposed algorithm in our simulations in  \cref{sec:simulation studies}.
	
	Next we update $\hat  \Z^{(k)}$ and $\hat  \Y^{(k)}$ as 
	\begin{eqnarray}
	&&\hat  \Z^{(k)}\label{eq:update Z} \in  
	 \arg \min_{\Z \in \Reals^{n \times p}} \frac{L_{k-1}}2 \left\| \Z -\hat  \Z^{(k-1)}  + L_{k-1}^{-1} \nabla \cL_{N} \left(\hat  \Y^{(k-1)} ;\W \right) \right\|_2^2 + \lambda \left\| \Z \right\|_*, \nonumber \\
	&  & \hat  \Y^{(k)}  \leftarrow \hat  \Z^{(k)} + \frac{k-1}{k+\rho -1 }\left( \hat  \Z^{(k)} - \hat   \Z^{(k-1)} \right). \label{eq:update Y}
	\end{eqnarray}
until convergence or a maximum number of iterations is reached. Here $\nabla \cL_{N}(\Z;\W)$ is the gradient function of $\cL_{N}(\Z;\W)$: 
\begin{equation} \nabla \cL_N (\Z;W)\triangleq  \left(\frac{\partial \cL_N}{\partial z_{ij}}\right)_{n\times p }=\left(\frac{N_i}{N}\cdot\frac{e^{z_{ij}}}{\dps \sum_{k=1}^p e^{z_{ik}}}  - \frac{W_{ij}}{N}\right)_{n \times p}  =  \begin{bmatrix}  \frac{N_1}{N}\nabla \cL_{N_1}(\Z_1;\W_1) \\ \vdots \\  \frac{N_n}{N}\nabla \cL_{N_n} (\Z_n;\W_n)
\end{bmatrix} \in \Reals^{n\times p}  
\end{equation}	
and $L_k$ is the reciprocal of step size in the $k$th iteration, which can be chosen by the following line search strategy: denote 
	$$
	\cF_L \left(\Z, \Y ;\W \right) = \cL_{N} (\Z;\W)  - \cL_{N} (\Y;\W) - \langle \Z - \Y, \nabla \cL_{N} (\Y) \rangle - 2^{-1} L \| \X - \Y \|_F^2,
	$$
	as the error of approximating $\cL_{N}(\Z;\W)$ by the second order Taylor expansion  with the second order coefficient as $L$.  In the $k$th iteration, we start with integer $n_k=1$ and let $L_k = \gamma^{n_k} L_{k-1}$ for certain scaling parameter $\gamma >1$, then repeatly increasing $n_k=1,2,\ldots$ until $\cF_{L_k}$$\left(\hat \X^{(k)}, \hat  \Y^{(k-1)} \right) \le 0$. In the optimization literature, $\displaystyle\frac{k-1}{k+\rho -1}$ and $\rho$ are, respectively,  referred to as the momentum term and friction parameter. We follow the suggestions by \cite{su2014differential,cao2020multisample} and set a high friction rate that $\rho \ge \frac92$. 
	
	More details of this algorithm with fixed tuning parameter $\lambda$ are summarized in  Algorithm  \ref{alg:nuclear clr} in  \cref{app:details of algorithms}, denoted as ${NuclearCLR}\left(\W, \lambda \right)$.
\subsection{An auto-tuning procedure}
\label{subsec:selection of tuning parameters}

Different from \cite{cao2020multisample}, we only have one tuning parameter $\lambda$ in \eqref{eq:nuclear norm estimation clr} and we search within a larger search region of $\lambda$  \citep{xu2013speedup,avron2012efficient}, that is, $\lambda$ is selected from 
$\left\{2^{-3}, 2^{-2}, \ldots, 2^{3},2^{4}\right\}$.  Similar to \cite{avron2012efficient}, our tuning parameter selection procedure is based on the criteria
\begin{equation}
R(\Z ) \doteq \frac{\cL_{\cN} \left(  \Z  \right)}{\left\| \Z  \right\|_{*}} + \frac{\left\| \Z  \right\|_{*}}{\cL_{\cN} \left(  \Z  \right)}, \label{eq:criteria for tuning parrameter selection}
\end{equation}
motivated  by the intuition that $\cL_{\cN}\left( \hat \Z (\lambda)\right)$ and $\lambda \left\| \hat \Z (\lambda)\right\|_*$ has to be of same magnitude \citep{avron2012efficient}; otherwise, one of $\cL_{\cN}(\Z)$ or nuclear-regularization $\lambda \| \Z \|_*$  dominates the other  in the optimization procedure: for example, if $\cL_{\cN}(\Z)$ is much larger than $\lambda \| \Z \|_*$, then the estimator might not be likely to have low-rank property since $\lambda \| \Z \|_*$  affects the optimization procedure in a limited way.

In first  step, we  initialize $\lambda^{(0)} = \cL_{N}\left(\hat  \Z^{(0)}\right)$ with $\hat  \Z^{(0)}$ in \eqref{eq:zero replacement}, and  for $l=0,1,2...$.  is to empirically set $\lambda \ge \cL_{\cN} (\hat \Z )  $; similar way of setting initial value for  the tuning parameter appears in \cite{avron2012efficient} as well. Theoretically speaking, this is consistent with the idea of $ \lambda$ having a lower bound \citep{shang2019tuning}; however, we are unable to establish the lower bounds  since we are unable to analytically derive duality of our objective function like \cite{shang2019tuning}.

In following iterative steps, we estimate $R\left(\hat \Z\left(\lambda^{(l)} \right) \right)$ and expects it decreases in first several iterations and stop when $ R \left( \hat \Z\left(\lambda^{(l)} \right) \right)$ starts increasing, that is, when $ R \left( \hat \Z\left(\lambda^{(l)} \right) \right)$ is close to its local minimum.  Similar to  \cite{avron2012efficient}, our search region is $\{ \gamma_{\lambda}^{l} \lambda^{(0) },l=0,1\ldots \}$ we set the  empirical scaling factor $\gamma_{\lambda}$ set to 1.5.

More details are summarized in Algorithm \ref{alg:auto-turning nuclear clr} in   Section \ref{app:details of algorithms}.

\ignore{
\begin{figure}
	\centering
	\begin{tabular}{c}
				\includegraphics[width=\textwidth]{../img/ratio_tuning_gamma_2_p50_no_outliers-64}
	\end{tabular}
	\caption{$R\left(\hat \Z (\lambda \right))$ versus $\lambda$ with $\gamma = 2, p = 50, n = 100 $ in simulation \ref{eq:data generating procedure V}. initial $\lambda$ is $\lambda^{(1)}$ in auto-tuning Algorithm \ref{alg:auto-turning nuclear clr}.}\label{fig:R Z lambda vs lambda}
\end{figure}	
}

\section{Theoretical Properties of the Proposed Estimator}
\label{sec:theoretical properties}

In this section, we investigate the theoretical properties of $\hat{Z}$ proposed in \eqref{eq:nuclear norm estimation clr} in \cref{sec:regularized estimation}; in particular, the upper bounds of the estimation  accuracy for \textsc{clr} matrix $\Z^*$ are provided in  Theorem \ref{thm:thm1} for the exact low-rank settings and Theorem \ref{thm:thm3} for the  approximate low-rank settings.  The following assumption appears in both settings to  ensure that  total number of the read counts are comparable across all the samples, which implies that the samples have similar read depths. 
\begin{assum} Denote $R_i$ for $i \in [n]$ which quantifies the proportion of the total count for the $i$th subject.
	Assume there exist constants $\alpha_{\R}$, $\beta_{\R}$ such that, for any $i \in [n]$, $\frac{\alpha_{\R}}n \le R_i \le \frac{\beta_{\R}}n$.
	\label{assum: proportion}
\end{assum}
This assumption also appears in \cite{cao2020multisample}. 
\subsection{Estimation bounds  under the exact low-rank matrix assumption}

The following theorem shows the estimation upper bound results  over a class of bounded low-rank \textsc{clr}  matrices:
\begin{equation}
\cB_0 (r) \triangleq \left\{ \Z \in \reals^{n\times p}:{\rm rank}(\Z)\le r \right\}.
\label{eq:bounded low-rank composition matrices}
\end{equation}

\begin{thm}
	\label{thm:thm1}
	Under Assumption \ref{assum: proportion} and $\Z^* \in \cB_0(r)$, with tuning parameter selected as 
	\begin{equation}
	\lambda = \delta  \frac{\beta_{\R} \vee \left(p \max_{i,j} X_{ij}^* \right) }{\left(p\min_{i,j} X_{ij}^{*} \right)^2 }\cdot \frac{\log (n+p)}N. \label{eq:tuning parameter}
	\end{equation}
Suppose that $N \ge (n+p)\log (n+p)$,	then 	there exists constant $C$ independent of $n,p,r$ such that
	{\allowdisplaybreaks	\begin{eqnarray*}
			&&	\frac1n \expc \left\| \hat \Z (\lambda) - \Z^* \right\|_F^2  \le C_1 (p)\cdot 
			\frac{r(n+p) \log (n+p)}{pN}   , 
		\end{eqnarray*}}
		with probability at least $1 - \frac3{n+p}$ where 
		$$C_1(p)=	\frac{C}{\min_{i,j}X_{ij}^* }
\cdot \frac{\left(\max_{i,j}X_{ij}^* \right)^2 \cdot \left\{ \beta_{\R} \vee \left(p \max_{i,j}X_{ij}^*\right) \right\}}{ \alpha_{\R} \left(\min_{i,j}X_{ij}^*\right)^3 }$$ 

\end{thm}
\ignore{In practive, initially we do not know 
$\frac{\beta_{\R} \vee \left(p \max_{i,j}\hat X_{ij}^* \right) }{\left(p\min_{i,j}\hat X_{ij}^* \right)^2}$ (a term inside tuning parameter $\lambda$) and we decide to use 
$\frac{\beta_{\R} \vee \left(p \max_{i,j}\hat X_{ij}^{(k)} \right) }{\left(p\min_{i,j}\hat X_{ij}^{(k)} \right)^2}$ instead and update $\lambda$ at each step..}

From Theorem \ref{thm:thm1},  by using the softmax transformation, we   can obtain an estimate of the compositional matrix $\X^*$, denoted as $\hat{\X}$.  The following Corollary \ref{cor:1} gives an estimation error bound  on KL divergence of estimation matrix 
\begin{cor}
	\label{cor:1}
Under Assumption \ref{assum: proportion} and $\Z^* \in \cB_0(r)$, with tuning parameter selected in \eqref{eq:tuning parameter}. Given a fixed constant $C_0 \ge \frac{6}{p \min_{i,j}X_{ij}^*\alpha_\R}$, if $(n+p)\log (n+p) \le N < C_0(n+p)^2 \log (n+p)$, we have
	\begin{eqnarray*}
		\frac1n \sum_{i=1}^nD_{KL}\left( \softmax(\Z^*), \softmax(\hat \Z (\lambda))\right)
		\le  C_3 (p)  \cdot \frac{r(n+p) \log (n+p)}{pN} .
\end{eqnarray*}
where $$C_3(p)=C \textcolor{black}{  \frac{\left[\max_{i,j}X_{ij}^* \right]^2 \cdot \left[  \beta_{\R} \vee \left( p\max_{i,j}X_{ij}^*\right) \right]}{ \alpha_{\R} \left[\min_{i,j}X_{ij}^* \right]^3 }}, C\text{ is independent of }n,p,N, \alpha_{\R}, \beta_{\R}.$$
\end{cor}

The techniques are related to recent work on matrix completion \citep{negahban2012restricted,cao2020multisample}, although our problem setup, method and sampling procedure are all distinct from matrix completion. We apply a peeling scheme by partitioning the set of all possible values of $\hat{\Z}$, and then derive estimation upper bounds for each of these subsets based on concentration inequalities.
\subsection{Estimator bounds  under approximate low-rank matrix assumption}
We now  consider the setting of approximately low-rank \textsc{clr} matrix with singular values of \textsc{clr} matrix $\Z^*$ belonging to an $\ell_q $ ball,
\begin{equation}
\cB_q\left( \rho_q\right) \triangleq \left\{ \Z \in \Reals^{n \times p }: \Z\1_p = \0_n, \sum_{i=1}^{n \wedge p} |\sigma_i(\Z)|^q \le \rho_q  \right\},
\label{eq:approximate low-rank ball}
\end{equation}
where $0 \le q \le 1$. In particular,if $q=0$ the $l_0$ ball $\cB_0(\rho_0)$ corresponds to the set of  bounded matrices with rank at most $\rho_0$. In general, we have the following upper bound result:

\begin{thm}
	\label{thm:thm3}
	Under Assumption \ref{assum: proportion} and   $\Z^*\in \cB_q(\rho_q)$, with tuning parameter selected by \eqref{eq:tuning parameter}, if $N \ge (n+p) \log (n+p), N = O\left( \rho_q p^{\frac{q}2} \frac{(n+p)^{2+ \frac{q}2}}{n^{\frac{q}2}} \log (n+p)\right)$ , the estimator $\hat \Z (\lambda)$ in \eqref{eq:nuclear norm estimation clr} satisfies:
	\begin{eqnarray*}
		&& \frac{1}n \expc \left\| \hat \Z (\lambda) - \Z^* \right\|_F^2 
		\le    C(n,p, q,\rho_q)\left\{ \frac{(n+p)\log (n+p)}{N} \right\}^{1 - \frac{q}2} 
	\end{eqnarray*}n+p
	with probability at least $1-\frac3{n+p}$ where 
	$$C(n,p, q,\rho_q)=\frac{C_1}{\min_{i,j}X_{ij}^* } 
	\cdot \frac{\rho_q p^{\frac{q}2}}{n^{\frac{q}2}} \left\{  \frac{\left[\max_{i,j}X_{ij}^* \right]^4 \cdot \left[  \beta_{\R} \vee \left(\max_{i,j}X_{ij}^*p\right) \right]}{ \alpha_{\R} \left[\min_{i,j}X_{ij}^*\right]^4 } \right\}^{1 - \frac{q}2}, $$
	and 
	$C_1$ is independent of $n,p,N,\alpha_{\R}, \beta_{\R}$.
	
\end{thm}

The rates of convergence of  Theorem \ref{thm:thm3} with $q=0$ and $\rho_0=r$ reduces to the exact low-rank setting in Theorem \ref{thm:thm1}.

\ignore{
	\subsection{Estimation of diversity index}
	\begin{corollary}
		Assume that the assumptions in Theorem \ref{thm:thm3} hold and the tuning parameter is selected by \eqref{eq:tuning parameter}. For any $\Z \in \cS$, there exists constant $C_2$ independent of $n,p,r$ such that
		\begin{eqnarray*}
			&& \frac{1}n \expc \left\| \hat \Z - \Z \right\|_F^2 \\
			& \le &  \frac{C_1}{\min_{i,j} X_{ij}^*} \\
			&& \cdot \frac{\rho_q p^{\frac{q}2}}{n^{\frac{q}2}} \left\{  \frac{\left[\max_{i,j}X_{ij}^* \right]^4 \cdot \left[  \beta_{\R} \vee \left(\max_{i,j}X_{ij}^*p\right) \right]}{ \alpha_{\R} \left[\min_{i,j}X_{ij}^*\right]^4 }  \frac{(n+p)\log (n+p)}{N} \right\}^{1 - \frac{q}2} \\
			&&	\frac{1}n \sum_{i=1}^n D_{KL}\left(clr^{-1} (\Z_i) ,clr^{-1} (\hat \Z_i)  \right)\\
			& \le &  C_1 \frac{\rho_q p^{\frac{q}2}}{n^{\frac{q}2}} \left\{  \frac{\left[\max_{i,j}X_{ij}^* \right]^4 \cdot \left[  \beta_{\R} \vee \left(\max_{i,j}X_{ij}^*p\right) \right]}{ \alpha_{\R} \left[\min_{i,j}X_{ij}^*\right]^4 }  \frac{(n+p)\log (n+p)}{N} \right\}^{1 - \frac{q}2} 
		\end{eqnarray*}
	\end{corollary}
}

\subsection{Estimation of singular subspace in the low-rank setting}
We assume that the true \textsc{clr} matrix $\Z^*$ with rank $r$  has the singular value decomposition 
$$\Z^*=\U D \V^T=\sum_{i=1}^r d_i \U_i \V_i^T, $$
where $\D=\diag \{d_1,\dots, d_r\}$ consists of the singular values of $\Z^*$  with $d_1>d_2>\cdots>d_r$; $\U=(\U_1,\cdots, \U_r)$ and $\V=(\V_1,\cdots, \V_r)$ are $\ell^2$ normalized left and right singular vectors. 
Given an estimate of the sample \textsc{clr} matrix $\hat{\Z}$,  it is often of interest to estimate its corresponding singular vectors by the corresponding singular value decomposition \citep{aitchison1983principal}, denote them as $\hat{\U}$ and $\hat{\V}$.  Similar to \cite{xia2018confidence,xia2019data}, we can provide an upper bound for singular subspace distance based on Theorem \ref{thm:thm1} as well as Weyl's lemma \ref{eq:Weyl's lemma} \citep{weyl1912asymptotische} and Davis-Kahan's $\sin \Theta$ theorem \citep{davis1963rotation,davis1965rotation,davis1970rotation}.

\begin{thm}\label{thm:principal subspace estimation}
	
	Under all the assumptions  in Theorem \ref{thm:thm3} and with tuning parameter selected as  in \eqref{eq:tuning parameter} and by further  imposing a lower bound on the $r$th largest singular value:
	\begin{equation}
	\sigma_{r+1}(\Z^* )  \le \frac12\sigma_{r}(\Z^* ),  \label{eq:lower bound of r th largest singular value}
	\end{equation}
	then for right singular vectors $\V_{\Z^*}, \V_{\hat \Z (\lambda)}$ and left singular vectors $\U_{\Z^*}, \U_{\hat \Z (\lambda)}$ we have 
	\begin{eqnarray*}
&& 		\left\| \sin \Theta \left( \V_{\hat {\Z}(\lambda)},  \V_{ \Z^*}\right) \right\|_{F}^2, \left\| \sin \Theta \left( \U_{\hat {\Z}(\lambda)},  \U_{ \Z^*}\right) \right\|_{F}^2  \\
& \le &  \frac{4\left\| \left( \hat \Z (\lambda)-  \Z^* \right) \V_{\Z^*} \right\|_{F}^2}{\sigma_r^2\left(\Z^* \right) } 
	 \le  C(n,p,q,\rho_q) \left\{    \frac{(n+p)\log (n+p)}{N} \right\}^{1 - \frac{q}2}.
	\end{eqnarray*}
	with probability at least $1 - \frac3{n+p}$ where 
	$$C(n,p,q,\rho_q)=\frac{2C_1 n}{\sigma_r^2\left(\Z^* \right)  \min_{i,j} X_{ij}^*} \\
	\cdot \frac{\rho_q p^{\frac{q}2}}{n^{\frac{q}2}} \left\{  \frac{\left[\max_{i,j}X_{ij}^* \right]^4 \cdot \left[  \beta_{\R} \vee \left(p \max_{i,j}X_{ij}^*\right) \right]}{ \alpha_{\R} \left[\min_{i,j}X_{ij}^*\right]^4 } \right\}^{1 - \frac{q}2}.$$
\end{thm}

%
%
%

\section{Simulation Studies}
\label{sec:simulation studies}

We now evaluate the numerical performances of the proposed estimator $\hat \Z^{\rm nuc}$  under exact low-rank settings  and approximate low-rank settings  by simulations in  \cref{subsec:low-rank simulation setting}, \cref{subsec:approximate low-rank simulation setting}.  To avoid confusion, estimator $\hat \Z^{\rm nuc}$ is different from estimator $\hat {\Z}(\lambda)$  mentioned in \cref{sec:theoretical properties}: the estimator $\hat \Z^{\rm nuc}$ utilizes auto-tuning procedure in \cref{subsec:selection of tuning parameters} but  $\hat {\Z}(\lambda)$s in \cref{sec:theoretical properties}  are for  fixed tuning parameter $\lambda$.

Data generating procedures are  divided into two steps:
\begin{enumerate}[(1)]
	\item generate \textsc{clr} matrix $\Z^*$;
	\item generate count matrix $\W$ according to Poisson-Multinomial model \eqref{eq:multinomial distribution}: generate $\displaystyle R_i = \frac{P_i}{\sum_{k=1}^n P_k}$ with $P_i \sim \text{Uniform}[1,10]$ for each individual $i \in [n]$. Based on $R_i$ and $\X^* = \softmax\left(\Z^*\right)$, the read counts are generated from the multinomial model, i.e. $W_i \sim \text{Mult}\left( n_i; X_i^* \right) $, where $N_i = \gamma n p R_i$, $\gamma = 1,2,3,4,5$. The sample size is $n=100$ and the number of taxa is  $p \in \{ 50, 100,150\}$ .
\end{enumerate}

The second step is the same for low-rank settings in \cref{subsec:low-rank simulation setting} and approximate low-rank settings in \cref{subsec:approximate low-rank simulation setting}. As a result, it suffices to focus on generating procedures of \textsc{clr} matrix  $\Z^*$.

\subsection{Low-rank simulation settings}
\label{subsec:low-rank simulation setting}

As we explained in the beginning of  \cref{sec:simulation studies}, it suffices to focus on generating procedure  of \textsc{clr} matrices $\Z^*$.  Let $\U  \in \Reals^{n \times r}$ with $U_{ij} \sim \cN(0,0.5 )$ and $r=20$. In order to simulate correlated compositional data arising from metagenomics, let $\V = 0.2\V^{(1)}  +\V^{(2)} \in \Reals^{p \times r}$, where 
\begin{equation}
\V^{(1)}_{ij} = \left\{ \begin{array}{cc}
1, & i = j; \\ v, & i \ne j\text{ with probability }q; \\ 1, & i \ne j\text{ with probability } 1-q.
\end{array}\right.,  \V^{(2)}_{ij} \sim \cN \left(0, 10^{-2}\right),
\label{eq:data generating procedure V}
\end{equation}
	where the choice of $(v,q)$ is specified in Table \ref{tab:simulation results my data generating procedure singular subspace} and such choice is the same for low-rank settings in Table \ref{tab:simulation results Z estimation (low-rank and approximate low-rank)}. Further steps of generating count matrices $\W^*$ are specified in the beginning of  \cref{sec:simulation studies}.

The results are summarized in Table \ref{tab:simulation results Z estimation (low-rank and approximate low-rank)} and Table \ref{tab:simulation results my data generating procedure singular subspace}. The proposed estimator $\hat{\Z}^{\rm nuc}$ outperforms the zero-replacement estimator $\hat{\Z}^{\rm zr}$ and singular value thresholding estimator $\hat{\Z}^{\rm svt}$ in almost all settings. In particular, the difference between th loss of $\hat{\Z}^{\rm nuc}$ and the other two becomes more significant for smaller $\gamma$, i.e., when the number of total read counts is small; and the settings with $p=50,100$ has more significant loss than the settings with $p=150$. Improvement of estimation errors measured by $\sin\Theta$ distance for  right singular subspaces in Table \ref{tab:simulation results my data generating procedure singular subspace} is  generally more modest  than improvement of \textsc{clr} matrices $\hat{\Z}$: for settings with $p=150$, we can hardly see improvement in Table \ref{tab:simulation results my data generating procedure singular subspace} although such an improvement is still significant for $\hat{\Z}$ in  low-rank settings (Table \ref{tab:simulation results Z estimation (low-rank and approximate low-rank)}).

\begin{figure}
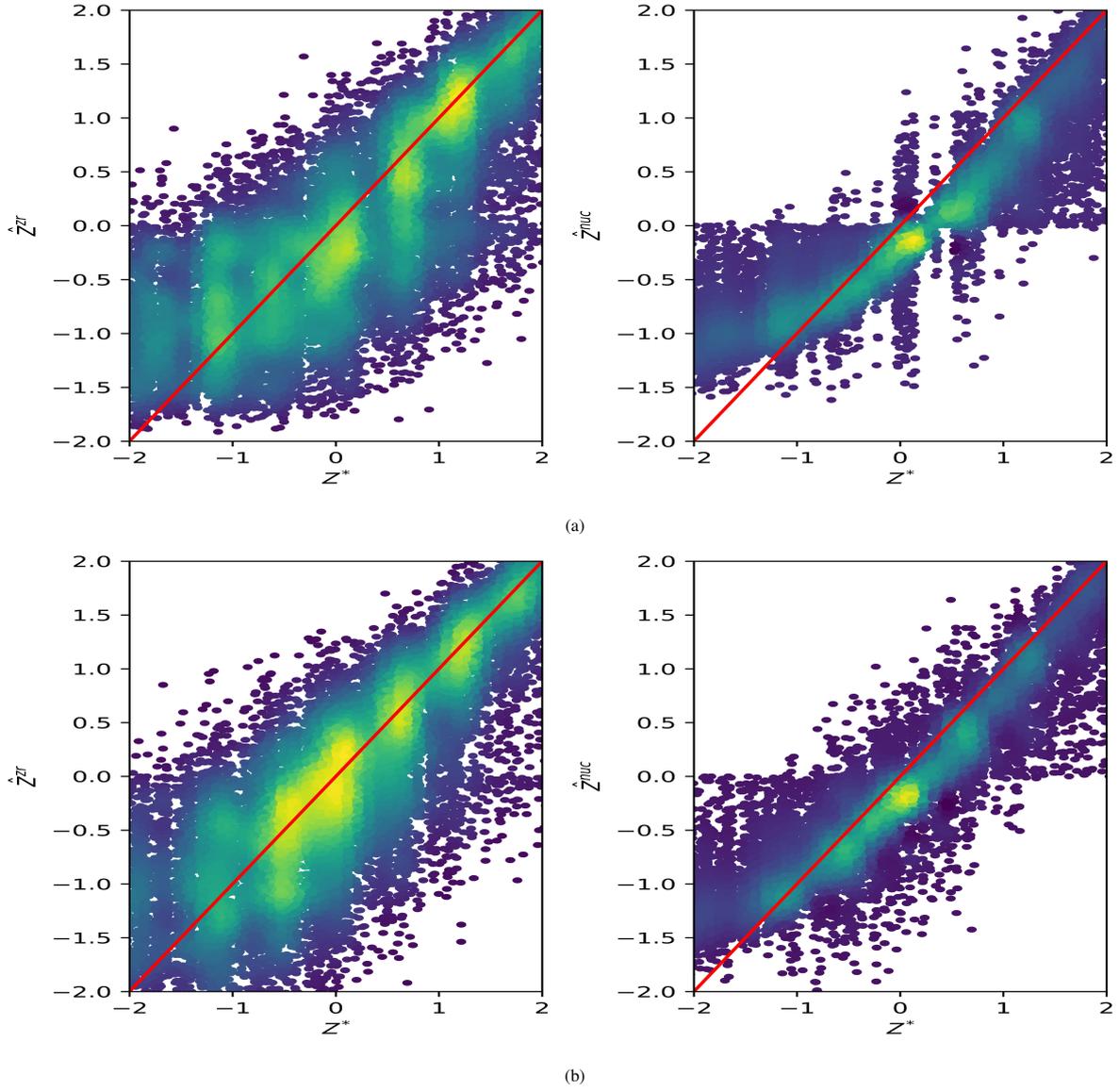

	\centering
			 \begin{subfigure}{\textwidth}
		\includegraphics[height=0.33\textheight,width=0.98\textwidth]{src/imgCh2/shrinkage_range-2-2_gamma_3_p50_no_outliers-200} 
									\caption{}
		\label{fig:1a shrinkage}
	\end{subfigure}\\
	\vspace{0.00mm}
			 \begin{subfigure}{\textwidth}
		\includegraphics[height=0.33\textheight,width=0.98\textwidth]{src/imgCh2/shrinkage_range-2-2_gamma_5_p50_no_outliers-200}
									\caption{}
		\label{fig:1b shrinkage}
	\end{subfigure}
	\caption{Scatter plots showing comparision of shrinkage of entries between $\hat{\Z}^{\rm zr}$ and $\hat{\Z}^{\rm nuc}$. Two settings are from the low-rank settings in Table \ref{tab:simulation results Z estimation (low-rank and approximate low-rank)}, Table \ref{tab:simulation results my data generating procedure singular subspace}. Two figures in \eqref{fig:1a shrinkage} correspond to the setting with $\gamma=3$, $p=50$ and  other two figures in \eqref{fig:1b shrinkage} correspond to the setting with $\gamma=5$, $p=50$.}
	\label{fig:Scatter plots showing shrinkage of entries}
\end{figure}

To further compare the resulting estimates, Fig.  \ref{fig:Scatter plots showing shrinkage of entries} shows two scatter plots comparing the true \textsc{clr} matrix $\Z^*$ and the estimated $\hat{\Z}$ for two low-rank settings in Table \ref{tab:simulation results Z estimation (low-rank and approximate low-rank)}. Although slightly biased due to the nuclear norm penalty in the estimation, it still greatly outperforms the commonly used zero-replacement estimator $\hat{\Z}^{\rm zr}$. 

\begin{table}
	\begin{center}
			\caption{Comparison of estimation errors measured by means of squared Frobenius norm error {\upshape $\left( \times 10^{-2} \right)$} of for $\hat \Z^{\rm nuc}$, $\hat \Z^{zr}$, $\hat \Z^{svt}$  for both exact and approximate low rank settings for various dimension $p$ and parameter $\gamma$.
		}
		\begin{tabular}{@{}lrrrrrrrrr@{}}
			\hline
			& \multicolumn{3}{c}{$p=50$} & \multicolumn{3}{c}{$p=100$}& \multicolumn{3}{c}{$p=150$}\\
			
			$\gamma$ & $\widehat{\Z}^{\rm nuc}$ & $\widehat{\Z}^{\rm zr}$ & $\widehat{\Z}^{\rm svt}$ & $\widehat{\Z}^{\rm nuc} $ & $\widehat{\Z}^{\rm zr}$ & $\widehat{\Z}^{\rm svt}$& $\widehat{\Z}^{\rm nuc}$ & $\widehat{\Z}^{\rm zr}$ & $\widehat{\Z }^{\rm svt}$\\
				\multicolumn{10}{c}{Low rank settings}\\
1&16.64&42.31&38.06&51.10&60.12&56.46&60.39&73.68&70.56\\
2&9.74&43.63&41.77&50.16&61.92&60.32&61.19&76.15&74.83\\
3&6.32&42.31&41.28&44.52&60.01&58.91&58.90&73.70&72.78\\
4&5.77&40.38&39.67&30.60&57.35&56.52&38.77&70.35&69.58\\
5&4.36&38.55&38.07&26.53&54.70&54.01&14.73&67.17&66.51\\
		\multicolumn{10}{c}{Approximate low-rank settings}\\			
1&31.77&43.43&42.33&57.97&61.54&60.37&74.02&75.76&74.91\\
2&28.49&41.53&40.96&56.98&58.99&58.22&71.85&72.61&71.93\\	
3&28.18&39.99&36.45&51.70&57.07&54.19&68.64&70.44&67.87\\
4&23.87&39.53&39.29&53.33&55.65&55.01&67.26&67.96&67.31\\
5&19.48&36.77&36.65&49.63&52.48&51.90&63.75&64.56&64.06\\		
\hline
		\end{tabular}
	\label{tab:simulation results Z estimation (low-rank and approximate low-rank)}
	\end{center}
	
\end{table}

\begin{table}
	\begin{center}
		\caption{Exact low-rank simulation settings: comparison of estimation errors measured by $\sin \Theta$ distance  for right singular subspaces for $\hat {\Z}^{\rm nuc}$, $\hat \Z^{\rm zr}$ and $\hat \Z^{\rm  svt}$ 		 Data generating procedure follows  \cref{sec:simulation studies} with $(v,q)=(-2,  0.5)$ in \eqref{eq:data generating procedure V} in Section \ref{subsec:low-rank simulation setting}.  This data generating procedure is the same for low-rank settings  in Table \ref{tab:simulation results Z estimation (low-rank and approximate low-rank)}.}
		\label{tab:simulation results my data generating procedure singular subspace}		\begin{tabular}{@{}lrrrrrrrrr@{}}
			\hline
			& \multicolumn{3}{c}{$p=50$} & \multicolumn{3}{c}{$p=100$}& \multicolumn{3}{c}{$p=150$}\\[5pt] 
			$\gamma$ & $\widehat{\V}^{\rm nuc}$ & $\widehat{\V}^{\rm zr}$ & $\widehat{\V}^{\rm svt}$ & $\widehat{\V}^{\rm nuc}$ & $\widehat{\V}^{\rm zr}$ & $\widehat{\V}^{\rm svt}$& $\widehat{\V}^{\rm nuc}$ & $\widehat{\V}^{\rm zr}$ & $\widehat{\V }^{\rm svt}$\\
			\hline
			\multicolumn{10}{c}{$\left\| \sin \Theta \right\|_F^2$  {\upshape $\left( \times 10^{-2} \right)$}  $\widehat{\V}^{\rm nuc}$, $\widehat{\V}^{\rm zr}$, $\widehat{\V}^{\rm svt} \in \Reals^{p \times 1}$}\\
1&66.25&173.47&166.67&109.80&183.52&183.81&187.68&188.40&187.12\\
2&62.88&172.27&173.38&79.59&180.76&182.27&185.81&186.86&186.74\\
			3&47.02&178.15&184.54&55.36&182.32&179.27&191.11&186.30&189.71\\
			4&45.37&170.72&182.15&50.32&181.99&181.97&194.59&187.25&175.22\\
			5&44.81&174.84&177.70&49.34&180.63&177.34&187.62&180.75&190.09\\
			\hline
			
	\multicolumn{10}{c}{$\left\| \sin \Theta \right\|_F^2$  {\upshape $\left( \times 10^{-2} \right)$ }  $\widehat{\V}^{\rm nuc}$, $\widehat{\V}^{\rm zr}$, $\widehat{\V}^{\rm svt} \in \Reals^{p \times 2} $ }\\
1&239.72&330.85&328.30&283.10&351.66&352.75&364.49&363.49&363.48\\
2&234.11&330.45&332.03&257.10&349.49&351.48&357.89&360.79&360.95\\
3&230.72&331.90&334.49&280.47&352.00&352.85&359.76&360.94&358.99\\
4&231.13&325.36&332.15&235.63&352.01&351.12&362.28&361.96&361.58\\
5&238.78&333.65&335.11&220.25&351.99&353.08&361.18&359.02&359.80\\
[5pt]
\hline			
\multicolumn{10}{c}{$\left\| \sin \Theta \right\|_F^2$  {\upshape $\left( \times 10^{-2} \right)$ }  $\widehat{\V}^{\rm nuc}$, $\widehat{\V}^{\rm zr}$, $\widehat{\V}^{\rm svt} \in \Reals^{p \times 3} $ }\\ 
1&394.14&472.05&472.84&450.04&514.22&512.28&531.68&528.68&530.43\\
2&388.94&475.94&476.22&423.53&511.51&512.25&526.91&528.88&529.81\\
3&384.03&480.26&472.65&447.31&515.79&512.42&527.39&528.44&527.09\\
4&383.81&473.70&479.97&402.53&516.39&515.56&527.69&531.43&528.90\\
5&387.86&478.80&480.26&389.71&518.93&512.66&529.37&529.42&531.04\\
[5pt] \hline 
\multicolumn{10}{c}{$\left\| \sin \Theta \right\|_F^2$  {\upshape $\left( \times 10^{-1} \right)$ }  $\widehat{\V}^{\rm nuc}$, $\widehat{\V}^{\rm zr}$, $\widehat{\V}^{\rm svt} \in \Reals^{p \times 20} $ }\\
1&173.80&176.30&175.25&241.95&246.37&245.46&275.43&274.77&275.43\\
2&173.88&176.29&176.62&241.17&246.40&245.66&275.38&275.11&275.12\\
3&173.64&176.11&177.47&242.27&246.36&246.43&274.27&275.88&274.85\\
4&173.47&176.56&177.36&240.32&245.73&245.93&274.43&275.42&275.01\\
5&173.88&176.20&177.02&239.37&246.40&245.10&275.14&274.79&276.08\\
\hline
\end{tabular}
\end{center}

\end{table}

\subsection{Approximate low-rank simulation settings}
\label{subsec:approximate low-rank simulation setting}
For the approximation low-rank settings, we try to identify a data generating procedure different from the exact low-rank settings in \cref{subsec:low-rank simulation setting}. As we explained in the beginning of  \cref{sec:simulation studies}, it suffices to focus on generating procedure  of \textsc{clr} matrices $\Z^*$.  Different from Section \cref{subsec:low-rank simulation setting}, we put $r = \min\{n,p\}$ but have $\Z^*= \tilde{\U}\D\tilde{\V }^T$ (instead of  $\Z^*= \U \V^T$) where 
\begin{enumerate}[(a)]
		\item $\tilde{\U } \in \Reals^{n \times r}$, $\tilde{\V } \in \Reals^{p \times r}$ are (column-wise $\ell_2$ normalized) right eigenvectors of $\U \in \Reals^{n \times r}$, $\V \in \Reals^{p \times \min\{n,p\}}$ to ensure diagonals of $\D$ can represent singular values of $\Z^*$.
	\item $\D = \diag\left\{i^{-2}, i=1,2,\ldots, \min\{n,p\} \right\}$ satisfy approximate low-rank assumption \eqref{eq:approximate low-rank ball} with $q=1$ (since $\displaystyle\sum_{i=1}^{\infty}i^{-2} = \frac{6}{\pi^2}< \infty$)
	\item 
	generate $\U \in \Reals^{n\times r }$ and $\V= 0,2\V^{(1)}+ \V^{(2)} \in \Reals^{p\times r }$ in the following way:
	\begin{equation*}
	\V^{(1)}_{ij} = \left\{ \begin{array}{cc}
	1, & i = j; \\ v, & i \ne j\text{ with probability }q; \\ 1, & i \ne j\text{ with probability } 1-q.
	\end{array}\right.,  \V^{(2)}_{ij} \sim \cN \left(0, 5\cdot 10^{-2}\right),
	\end{equation*}
	where the choice of $(v,q)$ is specified in Table \ref{tab:simulation results my data generating procedure singular subspace approximate low-rank} and this choice is the same for approximate low-rank settings in Table \ref{tab:simulation results Z estimation (low-rank and approximate low-rank)}. 

\end{enumerate}
 Further steps of generating count matrices $\W^*$ are specified in the beginning of  \cref{sec:simulation studies}.

\begin{table}
	\begin{center}
		\caption{Approximate low-rank simulation settings: comparison of estimation errors measured by $\sin \Theta$ distance  for right singular subspaces for $\hat {\Z}^{\rm nuc}$, $\hat \Z^{zr}$ and $\hat \Z^{svt}$ in the low rank model over 50 replications (we  run 50 replications since this is much slower than "low-rank" settings). Data generating procedure follows  \cref{sec:simulation studies} with $(v,q)=(-1,  0.5)$ in Section \ref{subsec:approximate low-rank simulation setting}. This data generating procedure is the same as for approximate low-rank settings  in Table \ref{tab:simulation results Z estimation (low-rank and approximate low-rank)}. }
		\label{tab:simulation results my data generating procedure singular subspace approximate low-rank}		\begin{tabular}{@{}lrrrrrrrrr@{}}
			\hline
			& \multicolumn{3}{c}{$p=50$} & \multicolumn{3}{c}{$p=100$}& \multicolumn{3}{c}{$p=150$}\\[5pt] 
			$\gamma$ & $\widehat{\V}^{\rm nuc}$ & $\widehat{\V}^{\rm zr}$ & $\widehat{\V}^{\rm svt}$ & $\widehat{\V}^{\rm nuc}$ & $\widehat{\V}^{\rm zr}$ & $\widehat{\V}^{\rm svt}$& $\widehat{\V}^{\rm nuc}$ & $\widehat{\V}^{\rm zr}$ & $\widehat{\V }^{\rm svt}$\\
			\hline
			\multicolumn{10}{c}{$\left\| \sin \Theta \right\|_F^2$  {\upshape $\left( \times 10^{-2} \right)$}  $\widehat{\V}^{\rm nuc}$, $\widehat{\V}^{\rm zr}$, $\widehat{\V}^{\rm svt} \in \Reals^{p \times 1}$}\\
1&170.96&172.75&173.88&177.80&186.74&170.86&187.33&184.54&188.53\\
2&180.30&175.97&182.29&179.70&182.22&180.70&188.56&186.28&187.65\\
3&175.88&178.53&169.75&180.56&177.31&180.01&188.24&186.55&187.24\\
4&180.29&180.16&175.62&183.38&184.16&185.00&186.80&187.23&188.73\\
5&176.49&180.79&175.45&187.91&182.89&182.60&184.18&188.16&189.31\\			\hline
			
			\multicolumn{10}{c}{$\left\| \sin \Theta \right\|_F^2$  {\upshape $\left( \times 10^{-2} \right)$ }  $\widehat{\V}^{\rm nuc}$, $\widehat{\V}^{\rm zr}$, $\widehat{\V}^{\rm svt} \in \Reals^{p \times 2} $ }\\
			1&333.34&331.12&339.27&351.64&347.54&345.49&361.45&362.41&358.81\\
			2&329.49&333.59&337.45&350.35&352.69&352.94&363.64&359.81&360.31\\
			3&338.77&343.03&333.04&353.88&349.34&350.01&364.20&357.38&365.85\\
			4&343.73&343.84&334.12&353.46&356.75&355.17&363.32&367.29&365.61\\
			5&331.76&334.34&342.92&357.04&353.80&357.61&358.72&361.38&368.84\\
						[5pt]
			\hline			
			\multicolumn{10}{c}{$\left\| \sin \Theta \right\|_F^2$  {\upshape $\left( \times 10^{-2} \right)$ }  $\widehat{\V}^{\rm nuc}$, $\widehat{\V}^{\rm zr}$, $\widehat{\V}^{\rm svt} \in \Reals^{p \times 3} $ }\\ 
1&464.15&488.80&472.40&512.67&507.97&507.28&527.99&522.48&525.37\\
2&478.23&480.28&477.62&510.23&506.52&512.77&526.78&531.88&528.96\\
3&468.29&491.70&472.51&513.62&508.42&517.00&527.73&530.37&530.40\\
4&488.29&486.52&480.52&505.49&517.41&515.26&536.24&536.76&531.08\\
5&477.98&472.42&484.11&522.21&518.09&512.62&524.05&531.84&536.62\\
			[5pt] \hline 
			\multicolumn{10}{c}{$\left\| \sin \Theta \right\|_F^2$  {\upshape $\left( \times 10^{-1} \right)$ }  $\widehat{\V}^{\rm nuc}$, $\widehat{\V}^{\rm zr}$, $\widehat{\V}^{\rm svt} \in \Reals^{p \times 20} $ }\\
1&174.99&175.57&175.69&243.59&243.28&244.26&276.34&276.72&275.47\\
2&176.66&176.13&178.83&246.08&246.03&244.15&274.31&273.95&275.32\\
3&179.43&176.85&176.68&246.70&244.33&244.78&275.04&275.88&274.45\\
4&176.39&176.49&177.03&247.01&247.38&247.75&275.19&276.05&275.95\\
5&177.19&173.74&175.30&244.59&247.42&245.51&274.97&275.03&275.58\\
			\hline
		\end{tabular}
	\end{center}
	
\end{table}

 We can see improvement in terms of estimation of $\hat {\Z}^{\rm nuc}$ in Table \ref{tab:simulation results Z estimation (low-rank and approximate low-rank)} but not much improvement $\hat {\V}^{\rm nuc}$ in Table \ref{tab:simulation results my data generating procedure singular subspace approximate low-rank}. While in exact low-rank settings, we have already seen that estimation of singular spaces are more difficult than estimating \textsc{clr} matrix $\Z^*$, here such phenomena appear again in the  approximate low-rank settings.

\section{Analysis of Real Datasets}
\label{sec:real data analysis}
We apply our \textsc{clr} matrix estimation algorithm in  \cref{sec:regularized estimation} to  \textcolor{black}{two real datasets}, the gut microbiome data set in a cohort of 98 individuals \citep{Wu2011} and the data set from the American Gut Project \citep{McDonalde00031-18}.  

\subsection{Gut Microbiome Dataset}
\label{subsec:Gut Microbiome Dataset}
The gut microbiome plays an important role in regulating metabolic functions and influences human health and disease \citep{methe2012framework,wu2016minimax}. \cite{Wu2011} reported a cohort gut microbiome data set that  includes  the counts of $87$ bacteria for $98$ healthy volunteers.  

\begin{figure}
	\includegraphics[height=0.33\textheight,width=\textwidth]{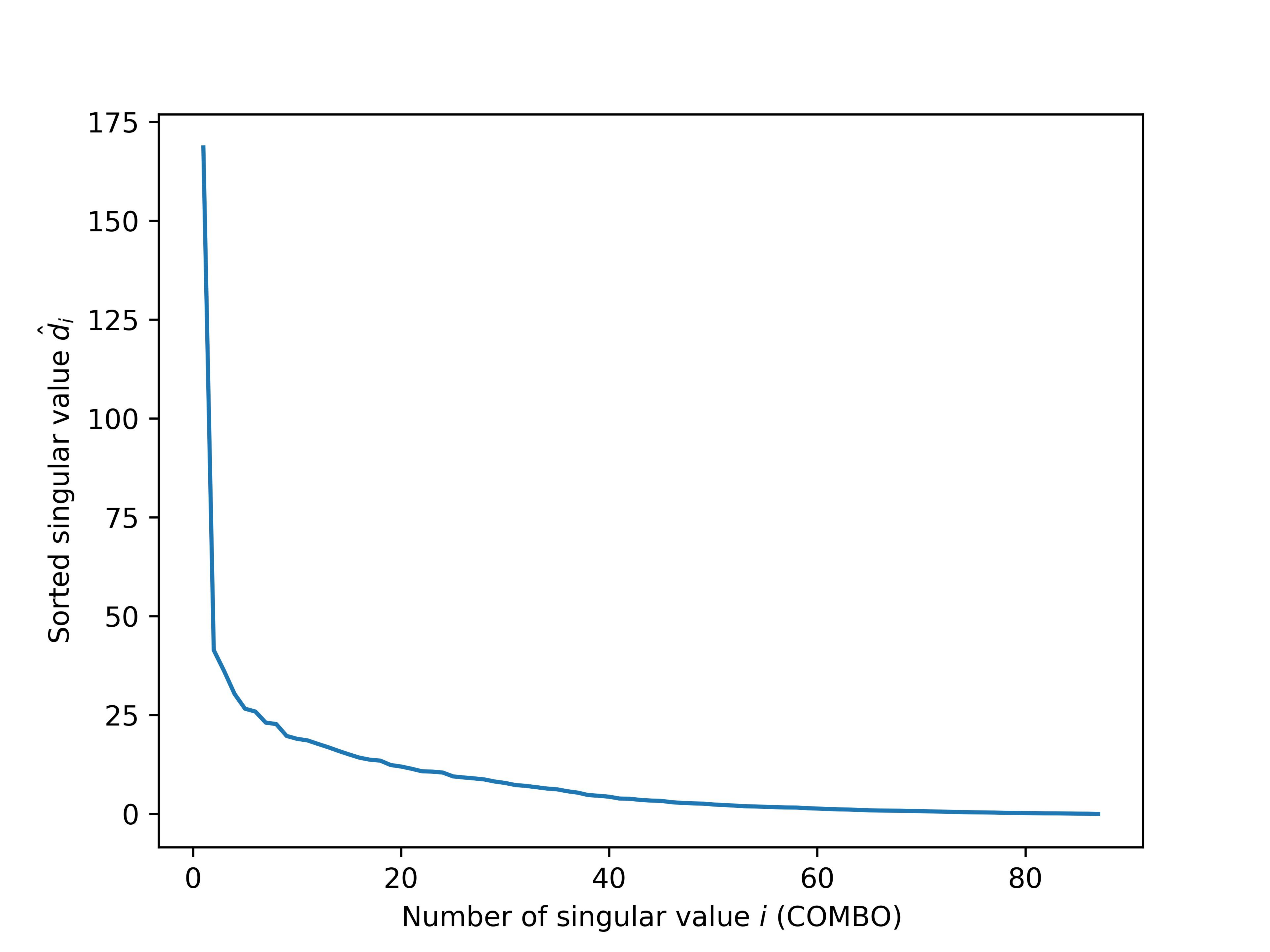}
	\caption{Analysis of Gut Microbime  dataset. The plot  shows the decay of singular values $ \hat d_i$ (versus $i$) based on
	the singular value decomposition of $\hat{\Z}^{\rm nuc}= \hat{\U}^{\rm nuc}\diag\{\hat{d}_1,\ldots, \hat{d}_{\min\{n,p\}}\}\left[\hat{\V}^{\rm nuc}\right]^T$ , indicating the low-rank structure of the compositional matrix.}\label{fig:a combo}
\end{figure}

  Fig. \ref{fig:a combo}  shows the decay singular values  $\hat{\Z}^{\rm zr}$ indicating an approximate low-rank \textsc{clr} matrix. 

\begin{figure}
	
	\centering
		\begin{subfigure}{\textwidth}
	\includegraphics[height=0.33\textheight,width=\textwidth]{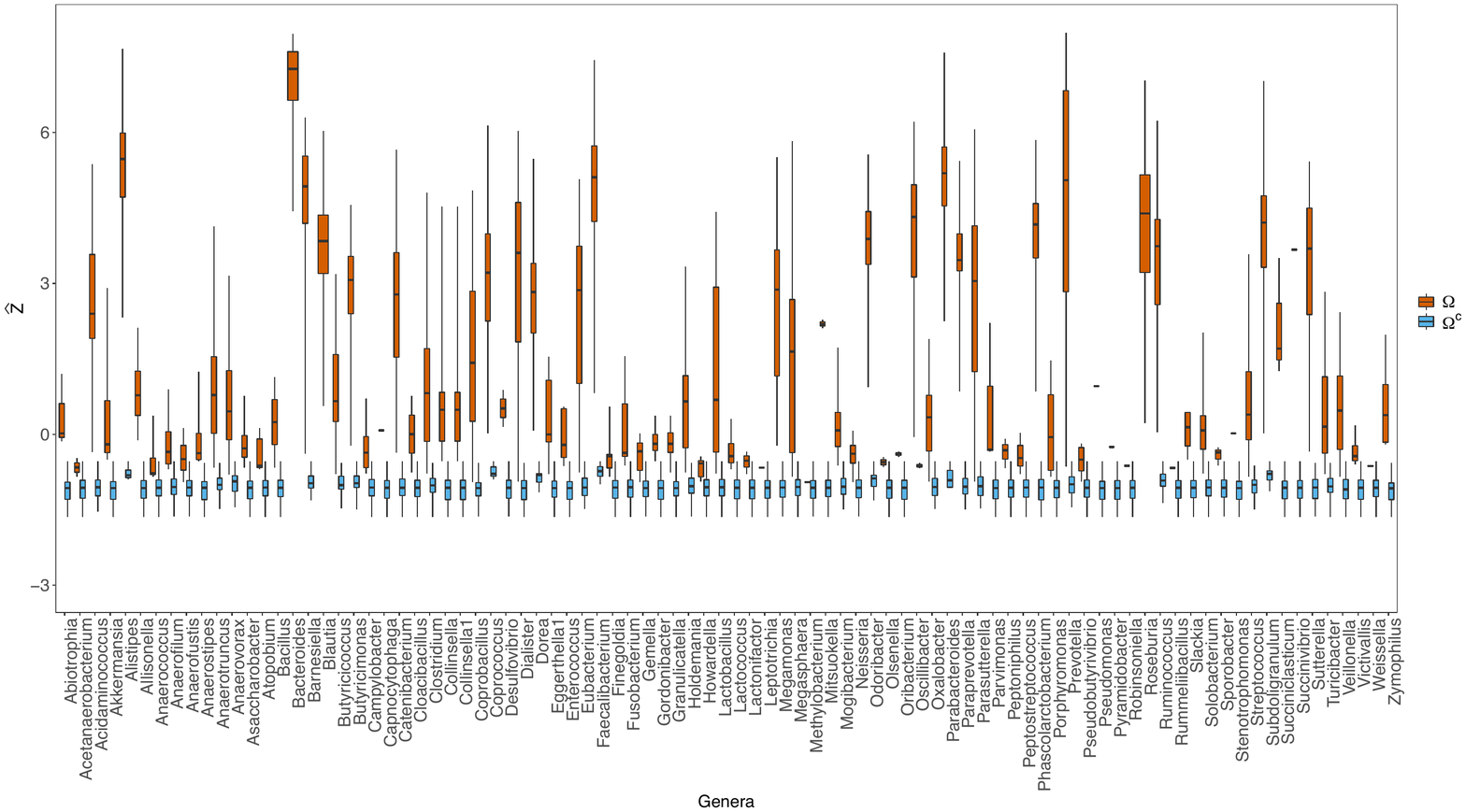}
	\caption{}
	\label{fig:a combo boxplot zr}
\end{subfigure}
	\begin{subfigure}{\textwidth}
	\includegraphics[height=0.33\textheight, width=\textwidth]{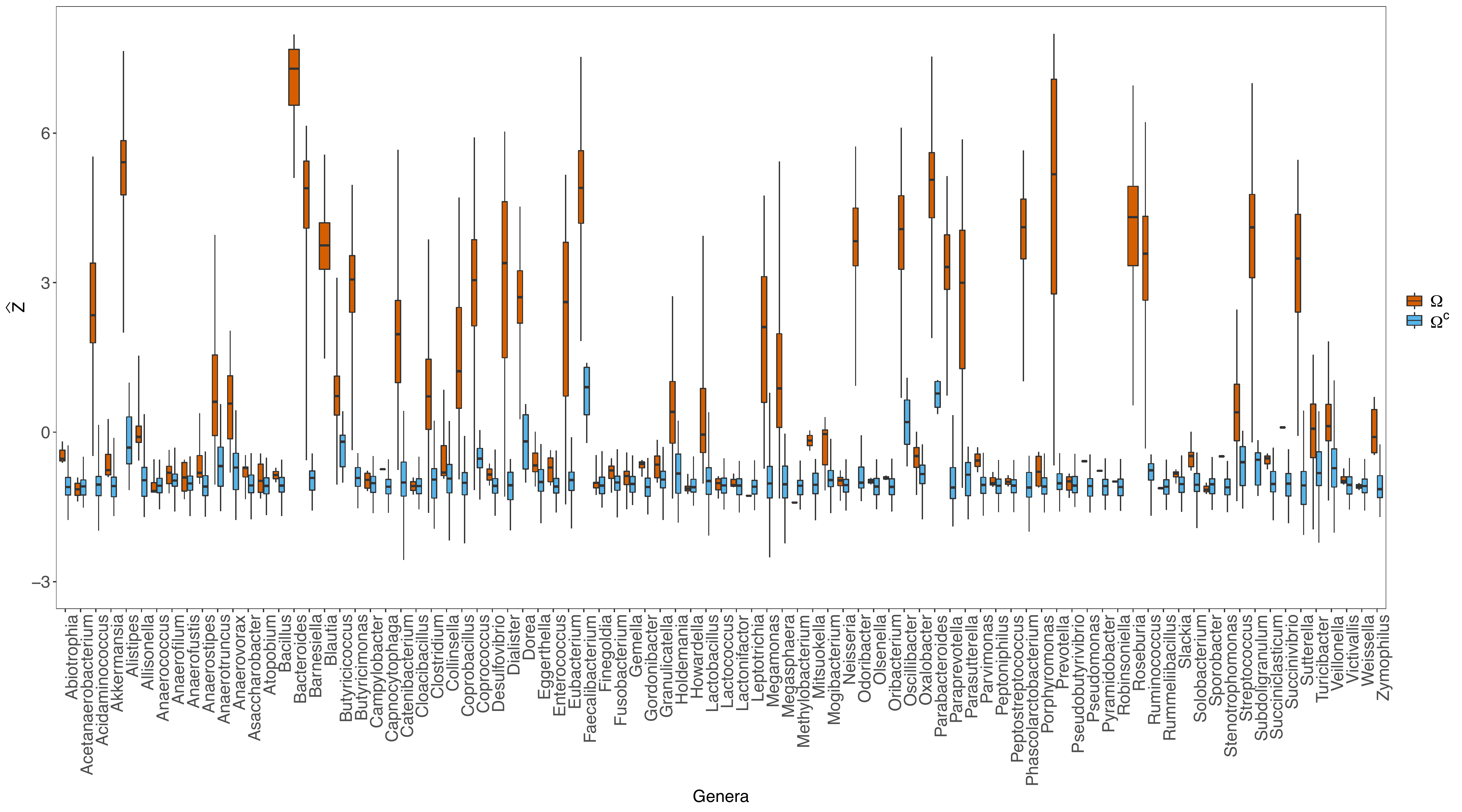}
		\caption{}
	\label{fig:b combo boxplot nuc}
\end{subfigure}

	\caption{Analysis of the gut microbiome dataset of \citep{Wu2011}.  Boxplots of the estimated center log-ratio transformation for the genera corresponding to non-zero observations ($\Omega$) and zero observations ($\Omega^c$).  Fig. \ref{fig:a combo boxplot zr}: the zero replacement estimator $\widehat{\Z}^{\rm zr}$.	 Fig. \ref{fig:b combo boxplot nuc}: the proposed estimator $\widehat {\Z}^{\text{auto}}$ with tuning parameter  set following  \cref{subsec:selection of tuning parameters} where the tuning parameter is auto-tuned as $\lambda = 2.15$.}
		\label{fig:boxplot Z combo}
\end{figure}

Fig. \ref{fig:boxplot Z combo} shows boxplots for \textsc{{clr}} matrices $\hat{\Z}^{\rm zr}$, $\hat{\Z}^{\rm nuc}$. To compare the results, define 
\begin{equation}\Omega = \{ (i,j): i=1,\ldots, n;j = 1,\ldots, p| W_{ij}> 0 \}
\label{eq:real dataset Omega}
\end{equation}
 and $\Omega^c$ as the support of the nonzero and zero entries in $\W$, respectively. Similar to \citep{cao2020multisample}, Fig. \eqref{fig:b combo boxplot nuc}  shows that the observed nonzero counts have an effect on estimating the \textsc{clr} matrix  of the genera that were observed as zeros. The estimated centered-log-ratio $\hat{Z}_{ij}^{\rm nuc}$ in $\Omega^c$ tends to shrink towards those in $\Omega$. In contrast, the zero-replacement estimator $\hat{\Z}^{\rm zr}$ in Fig. \eqref{fig:a combo boxplot zr} provides almost the same estimates for all the samples/taxa in $\Omega^c$ and $\{W_{ij}\}_{(i,j) \in \Omega}$, i.e. the observed nonzero counts have little effect on $\left\{ \hat{Z}_{ij}^{\rm zr}\right\}_{(i,j) \in \Omega^c}$.

\subsection{American Gut Project}
\begin{figure}

	\includegraphics[height=0.33\textheight,width=0.98\textwidth]{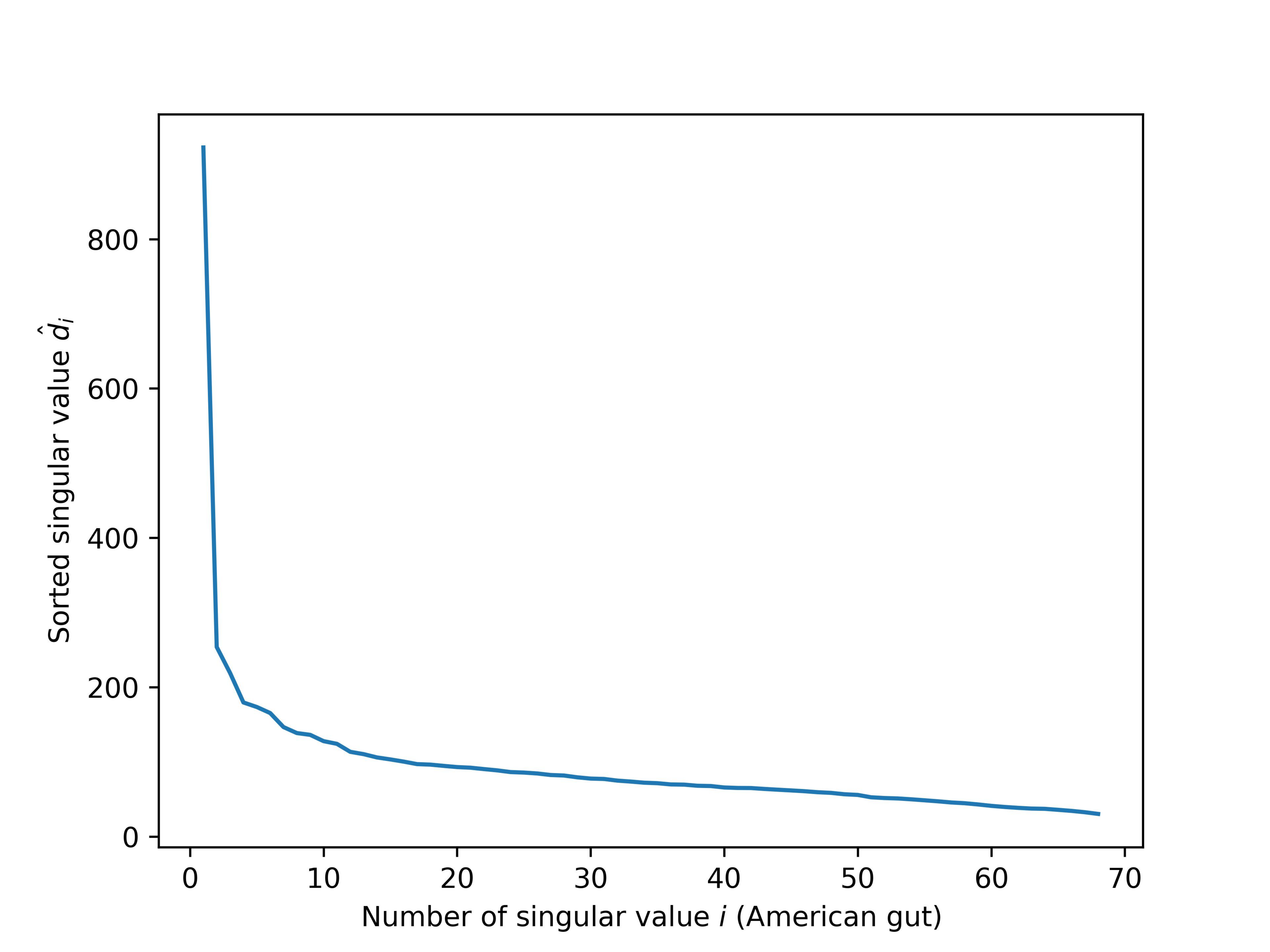}
	
	\caption{Analysis of the American Gut Project dataset: Fig. \eqref{fig:a american gut} shows decay of singular values $ \hat d_i$ (versus $i$) based on
	the singular value decomposition of $\hat{\Z}^{\rm nuc}= \hat{\U}^{\rm nuc}\diag\{\hat{d}_1,\ldots, \hat{d}_{\min\{n,p\}}\}\left[\hat{\V}^{\rm nuc)}\right]^T$ , indicating the low-rank structure of the compositional matrix. 
}	\label{fig:a american gut}

\end{figure}

The microbiome data of the American Gut Project \citep{McDonalde00031-18}  includes the counts of $70$ bacteria for $3,566$ individuals collected through an open platform for citizen science. Fig.  \ref{fig:a american gut} shows the decay of singular values  $\hat{\Z}^{\rm zr}$ indicating an approximate low-rank composition matrix. 

\begin{figure}
		
	\centering
		\begin{subfigure}{\textwidth}
	\includegraphics[height=0.33\textheight,width=\textwidth]{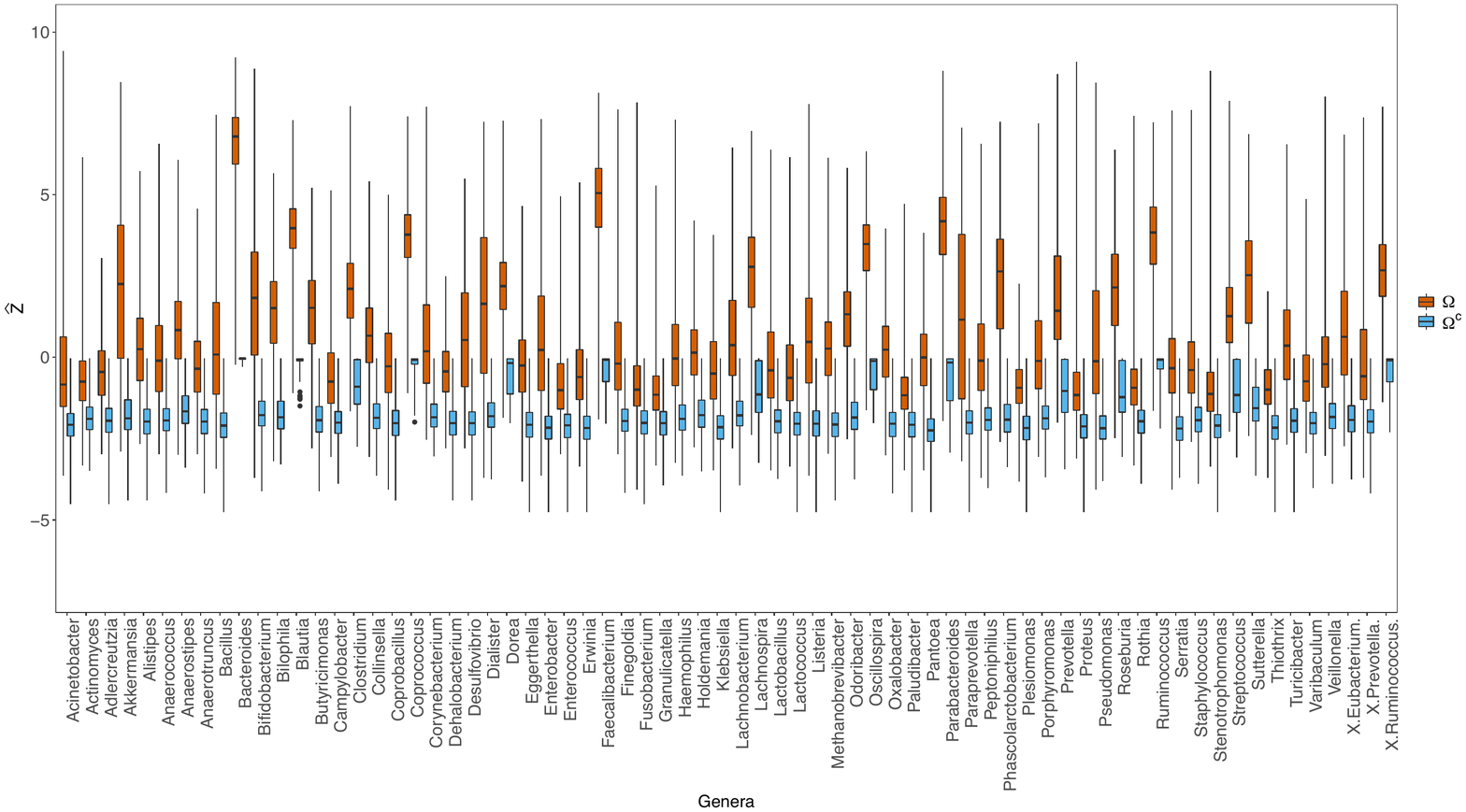}
	\caption{}
	\label{fig:a American gut boxplot zr}
\end{subfigure}
	\begin{subfigure}{\textwidth}
	\includegraphics[height=0.33\textheight,width=\textwidth]{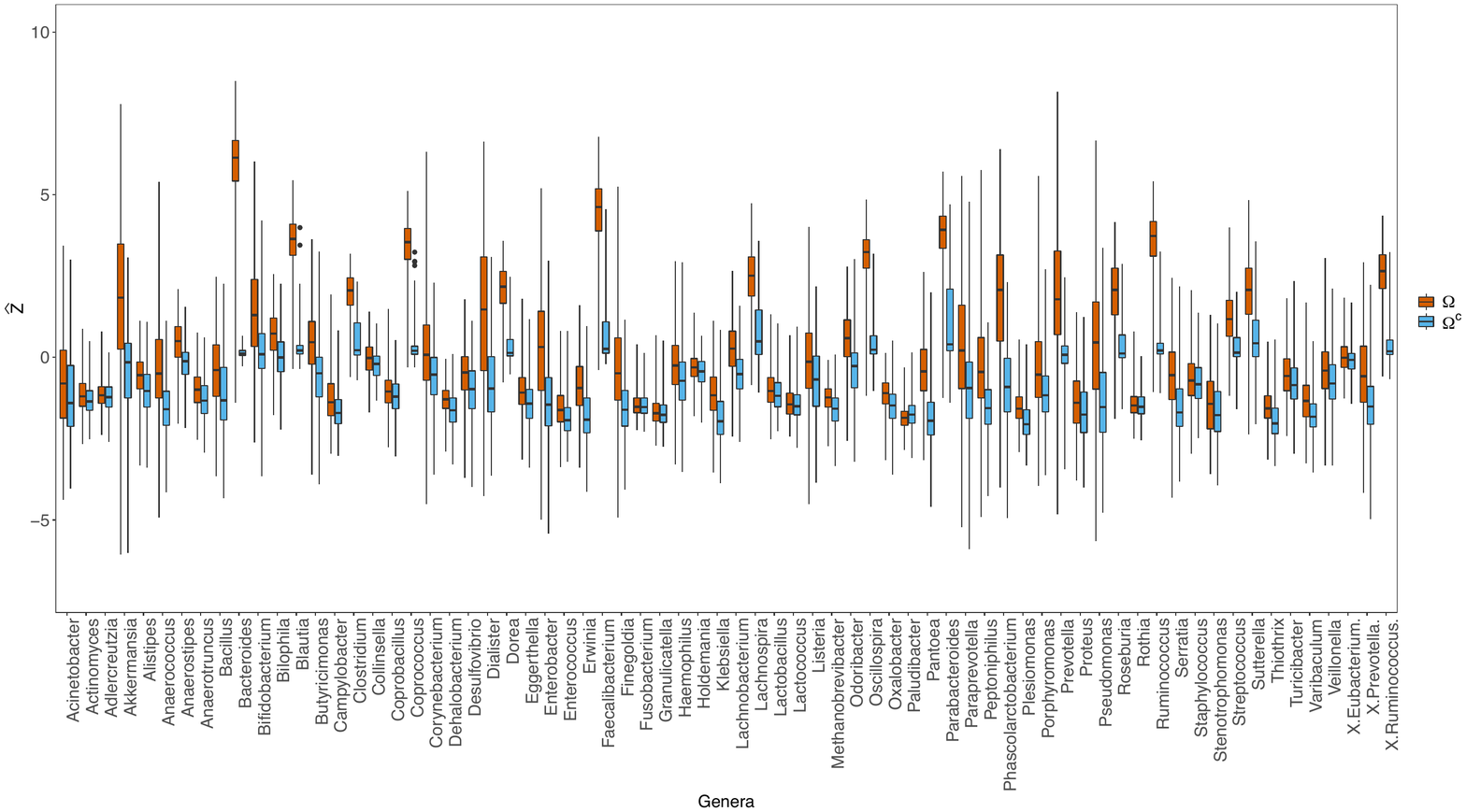}
		\caption{}
	\label{fig:b American gut boxplot nuc}
\end{subfigure}
\caption{Analysis of the  American Gut Project  dataset, showing the  boxplots of the estimated center log-ratio transformation of the compositions for the genera corresponding to non-zero observations ($\Omega$) and zero observations ($\Omega^c)$ in \textsc{combo} data set. 		 Top  panel: the zero replacement estimator $\widehat{\Z}^{\rm zr}$.	Bottom panel: the proposed estimator $\hat \Z^{\text{auto}}$ with the tuning parameter set following  \cref{subsec:selection of tuning parameters}, where the  tuning parameter is auto-tuned as $\lambda = 5.99$.}
\label{fig:American gut boxplot Z}
\end{figure}

Fig. \ref{fig:American gut boxplot Z} shows the boxplots of the estimated  \textsc{{clr}} matrices $\hat{\Z}^{\rm zr}$, $\hat{\Z}^{\rm nuc}$ ordered by their columns.  To compare the results,  Fig. \eqref{fig:b American gut boxplot nuc}  shows that the observed nonzero counts have much more effect on estimating the centered-log-ratio $Z_{ij}^*$ of the genera that were observed as zeros than $\hat{\Z}^{\rm zr}$ in Fig. \eqref{ fig:a American gut boxplot zr}. 

\section{Discussion}
Centroid-log-ratio transformation is  one of the most commonly used tranformations in compositonal data analysis. Traditionally the centroid-log-ratios are estimated from the compositional vectors. However, in many studies such as microbiome studies that motivated our method in this paper, the raw data are counts instead of the compositions, let alone centroid-log-ratios.  Treating the centroid-log-ratios as a parameter in Poisson-multinomial model for high dimensional count data, we have developed a nuclear-norm penalized maximum likelihood method for estimating the \textsc{clr} matrix of all the samples. The method effectively borrows information across the samples and taxa in order to achieve better estimation.  We  rarahave demonstrated this using simulations and analysis of the large real \textcolor{black}{datasets of  Gut Microbiome Dataset and  the American Gut Project}.
The method can be efficiently implemented using the generalized accelerated proximal gradient method.

\section*{Acknowledgment}
This research was supported by NIH grants GM129781 and GM123056.
\section*{\label{sec:supp}Supplementary Material}
Supplementary material related to this article can be found online.

\bibliographystyle{model5-names} 
\bibliography{src/compositional}

	\appendix

	
\subsection{Details of the algorithms}
\label{app:details of algorithms} 
This section provides more details of the algorithms described in  \cref{subsec:A generalized accelerated proximal gradient algorithm} and  \cref{subsec:selection of tuning parameters}.

Algorithm \ref{alg:nuclear clr} (denoted by ${NuclearCLR}\left(\W, \lambda \right)$ ) provides  more details than those appeared in  \cref{subsec:A generalized accelerated proximal gradient algorithm}, the generalized accelerated proximal gradient algorithm solving \eqref{eq:nuclear norm estimation clr} with fixed tuning parameter $\lambda$.

\begin{algorithm}[!h]
	\caption{${NuclearCLR}\left(\W, \lambda \right)$, the algorithm described in Section \ref{subsec:regularized estimation with fixed tuning parameter} for \eqref{eq:nuclear norm estimation clr}.
	}
	\label{alg:nuclear clr}
\begin{algorithmic}
		\REQUIRE Count matrix $\W \in \Naturals^{n \times p }$, tuning parameter $\lambda$. 
		
		Auxiliary parameters: scaling parameter $\gamma_L =1.5 $, initial step size $L_0$, friction rate $\rho = 5$, zero-criterion parameter $\epsilon_{\{\cF_{L} =0 \}} = 10^{-7}$; maximum number of iterations $K_{L,\max}=10^4$.
		\ENSURE Estimator of centroid log ratio matrix $\hat \Z^{(k)} \in \Reals^{n \times p}$.
		\STATE  Initialize $k=0$;
		\STATE \eqref{eq:zero replacement}: $\hat  Z_{ij}^{(0)}  = \hat Y_{ij}^{(0)} = \softmax\left( \frac{W_{ij} \wedge 0.5}{\sum_{j=1}^{p} \left(W_{ij}\wedge 0.5\right) } \right) + \epsilon$, where $\epsilon$ is the random perturbation with  $\epsilon  = \tilde{\epsilon }  \cdot  \left(\I_p - \frac1p \1_p\1_p\right)$ (and run $4,8,16, \ldots$ initializers due to this perturbation) ;
		\WHILE{$k \le K_{L,\max}$}
		\STATE $k \leftarrow k+1$;
		\STATE Determine $\hat \Z^{(k)}$, $\hat  \Y^{(k)}$ by \eqref{eq:update Z}, \eqref{eq:update Y} in the following way:
		\STATE Set $L_k \leftarrow L_0$;
		
		\WHILE{True }
		\STATE \eqref{eq:update Z} $\dps \hat  \Z^{(k)} \in \arg \min_{\Z \in \Reals^{n \times p}} \frac{L_{k-1}}2 \left\| \Z - \hat  \Z^{(k-1)}  + L_{k-1}^{-1} \nabla \cL_{N} \left(\hat  \Y^{(k-1)} ;\W \right) \right\|_2^2 + \lambda \left\| \Z \right\|_*$ by SVD and singular value thresholding;
		\STATE 	\eqref{eq:update Y}	$\hat  \Y^{(k)}  \leftarrow \hat  \Z^{(k)} + \frac{k-1}{k+\rho -1 }\left( \hat   \Z^{(k)} -  \hat \Z^{(k-1)} \right)$;
		\IF{$\cF_{L_k}\left( \hat \Z^{(k)}, \hat \Y^{(k)} ;\W\right) <0 $}
		\STATE break;
		\ENDIF
		\STATE $L_k \leftarrow \gamma_L L_k$.
		\ENDWHILE
		\IF{$\left| \cF_{L_k} \left(\hat  \Z^{(k)},  \hat  \Y^{(k)} ;\W \right) \right| \le \epsilon_{\{\cF_{L} =0 \}}$} 
		\STATE return $\hat  \Z^{(k)}$; (exit)
		\ENDIF
		\ENDWHILE
		\STATE return $\hat  \Z^{(k)}$; (exit)
	\end{algorithmic}
\end{algorithm}

Algorithm \ref{alg:auto-turning nuclear clr} provides  more details than those appeared in Section \ref{subsec:selection of tuning parameters} on how to process auto-tuning. The procedure is based on ${NuclearCLR}\left(\W, \lambda \right)$, that  is, Algorithm \ref{alg:nuclear clr}.

\begin{algorithm}[!h]
	\caption{Auto-tuning algorithm described in Section \ref{subsec:selection of tuning parameters}}
	\label{alg:auto-turning nuclear clr}
	\begin{algorithmic}
		\REQUIRE Count matrix $\W \in \Naturals^{n \times p }$. 
		
		Auxiliary parameters: maximum number of iterations $K_{\lambda,\max}=100$, stop criterion $\epsilon_{\lambda }= 10^{-3}$, scaling parameter $\gamma_{\lambda}= 1.2$.
		\ENSURE Estimator of centroid log ratio matrix $ \hat \Z^{\text{auto}} \in \Reals^{n \times p}$; selected tuning parameter $\lambda^{\text{auto}} $.
		\STATE  Initialize $k=0$;
		\STATE \eqref{eq:zero replacement}: $\hat Z_{ij}^{(0)}  =\hat Y_{ij}^{(0)} = \softmax\left( \frac{W_{ij} \wedge 0.5}{\sum_{j=1}^{p} \left(W_{ij}\wedge 0.5\right) } \right) $;
		\STATE  $\lambda^{(1)} \leftarrow \cL_{N}\left(\hat \Z^{(0)};\W\right)$;
		\STATE  $\lambda^{\text{auto}} \leftarrow \lambda^{(1)}$;
		\STATE criterion $R \leftarrow \frac{\cL_{N}\left(\hat \Z^{(0)};\W\right)}{\left\| \Z^{(0)}\right\|_{*}} + \frac{\left\| \hat \Z^{(0)}\right\|_{*}}{\cL_{N}\left(\hat \Z^{(0)};\W\right)}$;
		\WHILE{$k \le K_{\lambda , \max}$}
		\STATE $k \leftarrow k+1$;
		\STATE $ \hat\Z^{(k)} \leftarrow \mathfrak{NuclearCLR}\left(\W, \lambda^{(k)} \right)$, that is, Algorithm \ref{alg:nuclear clr};
		\STATE temporal $r \leftarrow  \frac{\cL_{N}\left(\hat \Z^{(k)};\W\right)}{\left\| \hat \Z^{(k)}\right\|_{*}} + \frac{\left\| \hat \Z^{(k)}\right\|_{*}}{\cL_{N}\left(\hat \Z^{(k)};\W\right)}$;
		\IF{$\frac{\left| R-r\right|}{R+r} \le \epsilon_{\lambda }$}
		\STATE $\lambda^{\text{auto}} \leftarrow \lambda^{(k)}$
		\STATE $\hat \Z^{\text{auto}} \leftarrow \mathfrak{NuclearCLR}\left(\W, \lambda^{\text{auto}} \right)$;
		\STATE return $ \hat \Z^{\text{auto}}$, $\lambda^{\text{auto}}$; (exit)
		\ENDIF
		
		\IF{$R >r$}
		\STATE $R \leftarrow r$;
		\STATE $\lambda^{\text{auto}} \leftarrow \lambda^{(k)}$;
		\STATE $\lambda^{(k+1)} \leftarrow \lambda^{(k)}*\gamma$;
		\ELSE
		\STATE $\lambda^{(k+1)} \leftarrow \sqrt{\lambda^{(k)} \lambda^{\text{auto}} }$
		
		\ENDIF
		\ENDWHILE
		\STATE $\hat \Z^{\text{auto}} \leftarrow \mathfrak{NuclearCLR}\left(\W, \lambda^{\text{auto}} \right)$;
		\STATE return $\hat \Z^{\text{auto}}$, $\lambda^{\text{auto}}$; (exit)
		
	\end{algorithmic}
\end{algorithm}

\subsection{Proof of Theorems}

For any integer $n>0$, we write $[n]=\{1,\ldots,n\}$ and denote $\e_i(n)$ as the canonical basis in $\reals^n$ with $i$th entry being  one and others being zero.

Before our derivation, we present Lemma \ref{lem:improved Davis-Kahan}, which is a consequence of Davis-Kahan $\sin \Theta$ theorem. While some classical forms are in \cite{stewart1990matrix,davis1970rotation}, we present Davis-Kahan $\sin \Theta$ theorem in the context  of our setting,  which is analogous to \cite{hsu2016COMS4772}:
\begin{lem}[Davis-Kahan $\sin \Theta$] \label{lem:improved Davis-Kahan}
	Denote singular value decomposition of symmetric matrix $\hat \A_m \in \Reals^{n \times n}$ as  $\hat \A_n  = \V_{\hat \A_n} \DiaEig_{\hat \A_n} \V_{\hat \A_n}^T + \V_{\hat \A_n}^{\perp} \DiaEig_{\hat \A_n}^{\perp} \left(\V_{\hat \A_n}^{\perp} \right)^T$, and similarly for  $\Z_n$. Suppose $ \left\| \DiaEig_{\hat \A_n}^{\perp}  \right\|_2  < \left\|  \DiaEig_{\A_n}^{-1}\right\|_2^{-1}$, where $  \left\| \DiaEig_{\A_n}^{- 1} \right\|_2^{-1 }    $ is the $r$th (absolutely) largest eigenvalue of $\A_n$, $     \left\| \DiaEig_{\hat \A_n} \right\|_2 $ is the $(r+ 1 )$th (absolutely) largest eigenvalue of $\hat \A_n$. Then for any unitarily-invariant norm $\| \cdot  \|_{\cU}$ (and we focus on $\|\cdot \|_{\cU} = \| \cdot \|_2, \| \cdot \|_{F}$),
	$$
	\left\| \left( \V_{\hat \A_n }^{\perp} \right)^T  \V_{\A_n}   \right\|_{\cU}  \le \frac{ \left\| \left(\hat \A_n - \A_n \right) \V_{ \A_n } \right\|_{\cU}  }{ \left\| \DiaEig_{\A_n}^{- 1} \right\|_2^{-1}  -  \left\| \DiaEig_{\hat \A_n}^{\perp} \right\|_2}= O_P\left( \frac{ \left\| \hat \A_n - \A_n \right\|_{\cU } }{ \sigma_r\left( \A_n \right)} \right). $$
\end{lem}

\subsubsection{Proof of Theorem \ref{thm:thm1} and Theorem \ref{thm:thm3}}
Theorem \ref{thm:thm1} and Theorem \ref{thm:thm3} can be considered as two special cases of following theorem:
\begin{thm}[Upper bounds]\label{thm:thm 4} With tuning parameter selected as \eqref{eq:tuning parameter}
	$$
	\lambda = \delta\frac{\beta_{\R} \vee \left(p \max_{i,j}\X_{ij}^*\right) }{\left(p\min_{i,j}\X_{ij}^*\right)^2 }\cdot \frac{\log (n+p)}N, 
	$$
	{\allowdisplaybreaks	\begin{eqnarray*}
			&&	\frac1n \left\| \Z^* -\hat \Z \right\|_F^2  \nonumber \\
			&\le & \frac{C}{\min_{i,j}\softmax(\Z_i^*)_j } \\
			&& \cdot \left\{ \left[\frac{\beta_{\R} \vee \left(\max_{i,j}\softmax(\Z_i^*)_jp\right) }{\left(\min_{i,j}\softmax(\Z_i^*)_jp\right)^2 }\cdot\frac{ \log (n+p)}{N} \right]^{\frac12} \sum_{i=r+1}^{n \wedge p} \sigma_i\left(\Z^*\right) \right. \nonumber \\
			&& \left. +\textcolor{black}{  \frac{\left[\max_{i,j}\softmax(\Z^*) \right]^2 \cdot \left[  \beta_{\R} \vee \left(\max_{i,j}\softmax(\Z^*)p\right) \right]}{ \alpha_{\R} \left[\min_{i,j}\softmax(\Z^*)\right]^3 } \cdot \frac{r(n+p) \log (n+p)}{pN}  }\right\}, \\
			&&			\frac{p}n \left\| \softmax\left( \Z_i^* \right) - \softmax \left( \hat \Z\right) \right\|_F^2 \nonumber \\
			&\le &   C \left\{ \left[\frac{\left[\max_{i,j}\softmax(\Z^*) \right]^4[\beta_{\R} \vee \left(\max_{i,j}\softmax(\Z_i^*)_jp\right) ]}{\left(\min_{i,j}\softmax(\Z_i^*)_jp\right)^4 }\cdot\frac{ \log (n+p)}{N} \right]^{\frac12} \sum_{i=r+1}^{n \wedge p} \sigma_i\left(\Z^*\right) \right. \nonumber \\
			&& \left. +\textcolor{black}{  \frac{\left[\max_{i,j}\softmax(\Z^*) \right]^4 \cdot \left[  \beta_{\R} \vee \left(\max_{i,j}\softmax(\Z^*)p\right) \right]}{ \alpha_{\R} \left[\min_{i,j}\softmax(\Z^*)\right]^4 } \cdot \frac{r(n+p) \log (n+p)}{pN}  }\right\} ,
		\end{eqnarray*}
		In addition, given a fixed constant $C_0 \ge \frac{6}{p \min_{i,j}\softmax(\Z_i^*)_j\alpha_\R}$, if $N < C_0(n+p)^2 \log (n+p)$, we have
		\begin{eqnarray*}
			&& \frac1n \sum_{i=1}^nD_{KL}\left(\Z^*, \hat \Z\right) \\
			&\le &   C \left\{ \left[\frac{\beta_{\R} \vee \left(\max_{i,j}\softmax(\Z_i^*)_jp\right) }{\left(\min_{i,j}\softmax(\Z_i^*)_jp\right)^2 }\cdot\frac{ \log (n+p)}{N} \right]^{\frac12} \sum_{i=r+1}^{n \wedge p} \sigma_i\left(\Z^*\right) \right. \nonumber \\
			&& \left. +\textcolor{black}{  \frac{\left[\max_{i,j}\softmax(\Z^*) \right]^2 \cdot \left[  \beta_{\R} \vee \left(\max_{i,j}\softmax(\Z^*)p\right) \right]}{ \alpha_{\R} \left[\min_{i,j}\softmax(\Z^*)\right]^3 } \cdot \frac{r(n+p) \log (n+p)}{pN}  }\right\} .
	\end{eqnarray*}}
	with probability at least $1 - \frac3{n+p}$.
\end{thm}
\begin{proof}
	Similar to \cite{cao2020multisample}, the count matrix $\W$ follows a multinomial distribution: $\W = \sum_{k=1}^N \E_k$ where $\E_k$ are independent and identically distributed copies of a Bernoulli random matrix $\E$ that  satisfies $$\Prob \left(\E = \e_{i}(n)\e_{j}(p)^T = \left[ \R \cdot \1_p^T\cdot \softmax \left(\Z^*\right) \right]_{ij}\right),$$
	where $\R$ is specified in  \cref{sec:theoretical properties}.
	
	Consequentially, 
	\begin{eqnarray}
	\cL_N(\Z) &=& - \frac1N \sum_{k=1}^N \log \langle \softmax(\Z), \E_k\rangle = - \frac1N \sum_{k=1}^N \log \langle \softmax(\Z), \E_k\rangle \nonumber \\
	& = & - \frac1N \sum_{k=1}^N \sum_{i=1}^n \log \langle \e_{i}(n)\e_{i}(n)^T \softmax(\Z), \E_k\rangle \nonumber \\
	& =  & - \frac1N\sum_{i=1}^n \sum_{k=1}^N  \log \langle \softmax(\Z_i)^T,  \ \E_k^T e_{i}(n)\rangle. \label{eq:MLE rewrite}
	\end{eqnarray}
	
	Any solution $\hat \Z$ to \eqref{eq:nuclear norm estimation clr} satisfies
	\begin{eqnarray}
	&& \cL_{N}\left(\hat \Z \right)  - \cL_{N}\left(\Z^* \right)\nonumber \\
	&=& \frac{1}{N} \sum_{i=1}^{n} \sum_{j=1}^{p } \left[ W_{ij} \log \left(\softmax \left( \Z_i^*\right)_j \right)  - W_{ij} \log \left(\softmax \left( \hat \Z_i\right)_j \right) \right] \nonumber \\ 
	&\le & \lambda \left( \left\| \Z^*\right\|_* -  \left\| \hat \Z \right\|_* \right) \label{eq:optimal inequality}.
	\end{eqnarray}
	
	Next we present following Lemmas to derive a lower bound for \eqref{eq:MLE rewrite}:
	\begin{lem}
		\label{lem:lem 1} Given the selected tuning parameter from Theorem \ref{thm:thm 4}, with probability at least $1 - \frac1{n+p}$, we have the following upper bound for $\left\| \Z^* - \hat \Z \right\|_*$:
		\begin{equation}
		\left\| \Z^* - \hat \Z \right\|_* \le 4 \sqrt{2r} \left\| \Z^* - \hat \Z \right\|_F + 4 \sum_{i=r+1}^{n \wedge p} \sigma_i \left( \Z^*\right).
		\label{eq:lem 1}
		\end{equation}
	\end{lem}
	\begin{lem}\label{lem:lem 2} If $\Z^*$ satisfies \eqref{eq:lem 1} as well as $\Z\1_p = \0_n$, we have 
		\begin{eqnarray}
		&&\frac1N \sum_{i=1}^n R_i D_{KL} \left(\softmax \left(\Z_i^* \right), \softmax\left(  \hat \Z_i\right) \right) \nonumber \\
		&\le & \cL_{N} \left(  \hat \Z \right)  - \cL_{N}\left(\Z^* \right) + C_2\frac{\beta_{\R} \vee \left(\max_{i,j}\softmax(\Z_i^*)_jp\right) }{\left(\min_{i,j}\softmax(\Z_i^*)_jp\right)^2 }\cdot \frac{ \log (n+p)}{N} \sum_{i=r+1}^{n \wedge p} \sigma_i\left(\Z^*\right)\nonumber  \\
		&  &	+\textcolor{black}{C_2 \frac{\left(\max_{i,j}\softmax(\Z_i^*)_jp \right)^2 \cdot \left[  \beta_{\R} \vee \left(\max_{i,j}\softmax(\Z_i^*)_jp\right) \right]}{ \alpha_{\R} \left(\min_{i,j}\softmax(\Z_i^*)_jp\right)^3 } \cdot \frac{r(n+p) \log (n+p)}{N}  }, \nonumber  
		\end{eqnarray}
		with probability proceeding $1 - \frac2{n+p}$.
	\end{lem}
	
	\begin{lem}\label{lem:lem 3} For any $\Z, \hat \Z\in \Reals^{n \times p}$ such that $\Z\1_p = \hat \Z\1_p = \0_n$, we have 
		\begin{equation}
		\min_{i,j} X_{ij}^*  \label{eq:equivalence distance} \le \frac{ \sum_{i=1}^n D_{KL} \left(\softmax \left(\Z_i^* \right), \softmax\left(  \hat \Z_i\right) \right)}{\left\|\Z^* - \hat \Z \right\|_F^2} \le  \max_{i,j}  X_{ij}^*.\nonumber 
		\end{equation}
	\end{lem}
	\begin{enumerate}
		\item First regime: $N < (n+p)\log (n+p)$,
		By applying Lemma \ref{lem:lem 1}, \ref{lem:lem 3}, we obtain the upper bound of $\|\Z^*\|_*-\|\hat \Z\|_*$ as 
		\begin{eqnarray*}
			\|\Z^*\|_*-\|\hat \Z\|_* & \le & \|\Z^*- \hat \Z \|_* \le 4\sqrt{2r} \|\hat \Z^*- \hat \Z\|_F + 4 \sum_{i=r+1}^{n \wedge p} \sigma_i \left( \Z^*\right) \\
			& \le & 	\sqrt {\frac{ \sum_{i=1}^n D_{KL} \left(\softmax \left(\Z_i^* \right), \softmax\left(  \Z_i^* \right) \right)}{ n \min_{i,j}   \softmax(\Z_i^*)_j}}+ 4 \sum_{i=r+1}^{n \wedge p} \sigma_i \left( \Z^*\right) .
		\end{eqnarray*}
		
		Therefore, combining \eqref{eq:optimal inequality}, Lemma \ref{lem:lem 1} and Lemma \ref{lem:lem 2} imply
		\begin{eqnarray*}
			&& \frac{\alpha_{\R}}{n} \sum_{i=1}^n D_{KL} \left(\softmax \left(\Z_i^* \right), \softmax\left(  \hat \Z_i \right) \right) \\
			& \le & \frac1N \sum_{i=1}^n R_i D_{KL} \left(\softmax \left(\Z_i^* \right), \softmax\left(  \hat \Z_i\right) \right)  \\ 
			&  \le & \lambda  \left[\sqrt {\frac{\dps \sum_{i=1}^n D_{KL} \left(\softmax \left(\Z_i^* \right), \softmax\left(  \hat \Z_i \right) \right)}{ n \min_{i,j}   \softmax(\Z_i^* )_j} }+ 4 \sum_{i=r+1}^{n \wedge p} \sigma_i \left( \Z^*\right)\right] \\
			&& +  \frac{\beta_{\R} \vee \left(\max_{i,j}\softmax(\Z_i^*)_jp\right) }{\left(\min_{i,j}\softmax(\Z_i^*)_jp\right)^2 }\cdot\frac{ \log (n+p)}{N}  \sum_{i=r+1}^{n \wedge p} \sigma_i\left(\Z^*\right) \\
			&  &	+\textcolor{black}{C_2 \frac{\left(\softmax(\Z^*)p \right)^2 \cdot \left[  \beta_{\R} \vee \left(\softmax(\Z^*)p\right) \right]}{ \alpha_{\R} \left(\softmax(\Z^*) p\right)^3 } \cdot \frac{r(n+p) \log (n+p)}{N}  },
		\end{eqnarray*}
		with probability at least $1-\frac3{n+p}$. The above formula can be treated as a quadratic inequality for $\dps \sum_{i=1}^n D_{KL} \left(\softmax \left(\Z_i^* \right), \softmax\left(  \hat \Z_i\right) \right)$. We plug in $\lambda$ in Theorem \ref{thm:thm 4}, and obtain
		\begin{eqnarray*}
			&&	\frac1n \sum_{i=1}^n D_{KL} \left(\softmax \left(\Z_i^* \right), \softmax\left(  \hat \Z_i\right) \right) \\
			&\le& C_3 \left\{ \frac{\beta_{\R} \vee \left(\max_{i,j}\softmax(\Z_i^*)_jp\right) }{\left(\min_{i,j}\softmax(\Z_i^*)_jp\right)^2 }\cdot\frac{ \log (n+p)}{N}  \sum_{i=r+1}^{n \wedge p} \sigma_i\left(\Z^*\right) \right. \\
			&& \left. +\textcolor{black}{  \frac{\left(\softmax(\Z^*)p \right)^2 \cdot \left[  \beta_{\R} \vee \left(\softmax(\Z^*)p\right) \right]}{ \alpha_{\R} \left(\softmax(\Z^*)p\right)^3 } \cdot \frac{r(n+p) \log (n+p)}{N}  }\right\}
		\end{eqnarray*}
		\item Second regime: $N>C_0 (n+p) \log (n+p)$. We denote $\triangle \doteq \hat \Z - \Z^*$. According to \eqref{eq:MLE rewrite} and Taylor expansion\ignore{ \eqref{eq:LN Taylor expansion} }, that is, there exists $t \in (0,1)$ such that 
		\begin{eqnarray*}
			&& \cL_{N} \left( \hat \Z    \right) - \cL_{N} \left(  \Z^*    \right) - \left\langle  \nabla \cL_{N} \left( \hat \Z    \right), \Delta \right\rangle \\
			& =  & \frac1N vec(\Delta)^T vec\left( \nabla vec\left( \nabla \cL_{N} \left( \hat \Z + t \Z^*    \right) \right)\right) vec( \Delta)  \\
			& =&    \frac1N vec(\Delta)^T\sum_{i=1}^n vec\left( \nabla vec\left( \nabla \cL_{N_i} \left(t \hat \Z + (1-t) \Z^*  \right) \right)\right) vec( \Delta), 
		\end{eqnarray*}
	\end{enumerate}
\end{proof}

\subsection{Proof of Lemmas}
\subsubsection{Proof of Lemma \ref{lem:lem 1}}
For notational simplicity, we denote $\Delta \triangleq \hat\Z - \Z^* \in \Reals^{n\times p}$, and 
{\allowdisplaybreaks
	\begin{eqnarray}\nonumber \nabla \cL_N (\Z)& \triangleq & \left(\frac{\partial \cL_N}{\partial z_{ij}}\right)_{n\times p }=\left(\frac{N_i}{N}\cdot\frac{e^{z_{ij}}}{\dps \sum_{k=1}^p e^{z_{ik}}}  - \frac{W_{ij}}{N}\right)_{n \times p} \in \Reals^{n\times p} \\
	& = &  \begin{bmatrix}  \frac{N_1}{N}\nabla \cL_{N_1}(\Z_1) \\ \vdots \\  \frac{N_n}{N}\nabla \cL_{N_n} (\Z_n)
	\end{bmatrix}  \nonumber \\
	&  \xlongequal{\text{\eqref{eq:MLE rewrite}}} &- \frac1N  \sum_{k=1}^N \begin{bmatrix} \frac{\e_{1}(n)^T\E_k \cdot\nabla {clr}^{-1}(\Z_1) }{\langle \softmax(\Z_1)^T,\E_k^T\e_{1}(n) \rangle} \\ \vdots \\ \frac{\e_{n}(n)^T\E_k \cdot\nabla {clr}^{-1}(\Z_n) }{\langle \softmax(\Z_n)^T, \E_k^T\e_{n}(n) \rangle} 
	\end{bmatrix} = \begin{bmatrix}
	\frac{\e_{i}(n)^T\E_k \cdot\nabla {clr}^{-1}(\Z_i) }{\langle \softmax(\Z_i)^T, \E_k^T \e_{i}(n) \rangle}\end{bmatrix}_{i\in [n]} , \label{eq:nabla of MLE rewrite}
	\end{eqnarray} }

Denote $vec(\Delta), vec (\nabla \cL_N ) \in \Reals^{np}$ vectorized forms of corresponding matrices.
According to Taylor expansion, 
\begin{eqnarray}
	&& \cL_{N} \left( \hat \Z    \right) - \cL_{N} \left(  \Z^*    \right) - \left\langle  \nabla \cL_{N} \left( \hat \Z    \right), \Delta \right\rangle \nonumber \\
	& =  & \frac1N vec(\Delta)^T vec\left( \nabla vec\left( \nabla \cL_{N} \left( \hat \Z + t \Z^*    \right) \right)\right) vec( \Delta) \nonumber \\
	& =&    \frac1N vec(\Delta)^T\sum_{i=1}^n vec\left( \nabla vec\left( \nabla \cL_{N_i} \left( \hat \Z + t \Z^*  \right) \right)\right) vec( \Delta), \label{eq:LN Taylor expansion}
\end{eqnarray}
and furtherly we obtain
\begin{eqnarray*}
	0 &\le & \frac1N vec(\Delta)^T\sum_{i=1}^n vec\left( \nabla vec\left( \nabla \cL_{N_i} \left( \hat \Z + t \Z^* \right) \right)\right) vec( \Delta) \\
	&=& \cL_{N} \left( \hat \Z    \right) - \cL_{N} \left(  \Z^*    \right) - \left\langle  \nabla \cL_{N} \left( \hat \Z    \right), \Delta \right\rangle \\
	& \le  &  - \left\langle  \nabla \cL_{N} \left( \hat \Z    \right), \Delta \right\rangle  + \lambda \left( \left\| \Z^*\right\|_* -  \left\| \hat \Z \right\|_* \right) \\
	& \le & \left\| \nabla \cL_{N} \left( \hat \Z    \right)\right\|_2 \cdot \left\|\Delta \right\|_*+ \lambda \left( \left\| \Z^*\right\|_* -  \left\| \hat \Z \right\|_* \right) 
\end{eqnarray*}

To further upper bound the nuclear norm $\| \hat \Z - \Z^*\|_* $, we state two technical results:
\begin{lem}
	\label{lem:lem 4}
	With probability at least $1 - \frac1{n+p}$, we have 
	$$
	\left\| \nabla \cL_{N} (\Z^*)\right\|_2 \le \left[\frac{M}3 + \sqrt{\frac{M^2}9 + \frac{\sigma^2}{ \log (n+p)}}\right]\frac{2 \log (n+p)}N \le c \frac{\log (n+p)}N,
	$$
	$\dps \sigma^2 = 1- \sum_{i=1}^n\nu_i \left\| \softmax\left(\Z_i^*\right) \right\|_2^2 $, 
	$$M = \sqrt{1 + \max_{i\in [n]}\left[  \left\|{clr}^{-1}(\Z_i^*)\right\|_2^2 - 2 \min_{j \in [p]}{clr}^{-1}(\Z_i^*)_j \right]}.$$
\end{lem}

Based on Lemma \ref{lem:lem 4}, with probability proceeding $1 - \frac1{n+p}$, the selected tuning parameter $\lambda \ge 2 \left\| \nabla \cL_{N} (\Z^*)\right\|_2$. 

According to Lemma 1 (B.2) in \cite{negahban2012restricted} as well as Lemma 5 in \cite{cao2020multisample}, \textcolor{black}{we obtain Lemma \ref{lem:lem 1}}.

\textbf{Proof of Lemma \ref{lem:lem 2}}

For notational simplicity, we denote
\begin{eqnarray*}
\eta & \triangleq & n \log \frac{\max_{i,j} X_{ij}^* }{\min_{i,j} X_{ij}^*}\left[ \frac{512 \log (n+p)}{ \log 4 \alpha_{\R}^2N} \right]^{\frac12 } , \\
	D_{\R} \left(\softmax\left(\Z^*\right),\softmax(\Z)\right)  & \triangleq  & \sum_{i=1}^n R_i D_{KL}\left(\softmax\left(\Z_i^*\right),\softmax(\Z_i)\right).
\end{eqnarray*}
The main lines of this proof are in the same spirit as Lemma 2 in \cite{cao2020multisample} as well as Lemma 3 in \cite{negahban2012restricted}. We use a peeling argument to prove the probability of the following "bad" event is small:
\begin{eqnarray*}
	&& \cB \triangleq\\
	&&  \left\{  \Z \in \Reals^{n \times p }:\left|  {\tiny\frac1N \sum_{k=1}^N \log \langle \softmax(\Z) - \softmax\left(\Z^*\right), \E_k\rangle - D_{\R} \left(\softmax\left(\Z_i^*\right),\softmax(\Z_i)\right) }\right| \right.\\
	&& \left. \ge D_{\R} \left(\softmax\left(\Z_i^*\right),\softmax(\Z_i)\right) +E(n,p,r) \right\},
\end{eqnarray*}
where $E (n,p,r)$ is defined by
{\allowdisplaybreaks
	\begin{eqnarray}
&& E(n,p,r) \triangleq \label{eq:E(n,p,r)} \\
 &&  \left[ \frac{\left[ (\beta_{\R}/ n) \vee \max_{i,j} X_{ij}^* \right]\log (n+p)}{N}+ \frac{\log (n+p)}{N}\right] \nonumber \\
&& \cdot \left\{ \frac{2048 \left[\max_{i,j} \softmax (\Z_i^*)_j\right]^2 nr}{\left[\min_{i,j} \softmax (\Z_i^*)_j\right]^3 \alpha_{\R}}  \left(\frac{ \max_{i,j} X_{ij}^*  }{  \min_{i,j} X_{ij}^*}\right)^2 \right. \nonumber \\
&&  \cdot \left[ \frac{\left[ (\beta_{\R}/ n) \vee \max_{i,j} X_{ij}^* \right]\log (n+p)}{N} \frac{\log (n+p)}{N}\right] \\
&& +  \left.\frac{16}{\min_{i,j} \softmax (\Z_i^*)_j}\sum_{i=r+1}^{n \wedge p} \sigma_i\left(\Z^*\right) \right\}, \nonumber
\end{eqnarray}
}
We separate the constraint set $\{\Z: \Z\1_p = \0_n\}$ into pieces and focus on a sequence of small sets:
$$
\cC_l \triangleq \left\{ \Z: \Z\1_p = \0_n, 2^{l-1}\eta \le \sum_{i=1}^n D_{KL}\left(\softmax\left(\Z_i^*\right),\softmax(\Z_i)\right) \le 2^l\eta \right\}, l =1,2,3,\ldots
$$

Notice 
$$
	D_{\R} \left(\softmax\left(\Z^*\right),\softmax(\Z)\right)  \ge \frac{\alpha_{\R}}{n}\sum_{i=1}^n  D_{KL}\left(\softmax\left(\Z_i^*\right),\softmax(\Z_i)\right),
$$
it suffices to estimate the probability of the following events and then apply the union bound,
\begin{eqnarray*}
	&& \cB_l \triangleq \\
	&& \left\{ \exists \Z:\left|  \frac1N \sum_{k=1}^N \log \langle \softmax(\Z) - \softmax\left(\Z^*\right), \E_k\rangle - D_{\R} \left(\softmax\left(\Z_i^*\right),\softmax(\Z_i)\right)\right| \right.\\
	&& \left. \ge \frac{2^l \eta \alpha_{\R}}{4n} +E(n,p,r) , \Z\1_p=\0_n \right\},
\end{eqnarray*}
Since $\cC_l \subset \cD (2^l\eta) \triangleq \left\{ \Z: \Z\1_p=\0_n, D_{\R} \left(\softmax\left(\Z_i^*\right),\softmax(\Z_i)\right) \le 2^l\eta \right\}$ we can establish the upper bound of the probability of event $\cB$ by using the union bound, the fact that $x \ge \log x $ and Lemma \ref{lem:lem 8}:
\begin{eqnarray*}
	\Prob \left( \cB \right) & \le & \sum_{l = 1}^\infty \Prob \left( \cD(2^l \eta) \cap \cB_l \right) \\
	& \le  &  \sum_{l = 1}^{\infty}\exp\left[ - \frac{\alpha_{\R}^2 N \eta^2 4^l}{512 \left(n \log \frac{\max_{i,j} \softmax\left(\Z_i^* \right)_j }{\min_{i,j} \softmax\left(\Z_i^*\right)_j }\right)^2 } \right] \\
	& \le &  \sum_{l = 1}^{\infty}\exp\left[ - \frac{\alpha_{\R}^2 N \eta^2 l\log 4 }{512 \left(n \log \frac{\max_{i,j} \softmax\left(\Z_i^* \right)_j }{\min_{i,j} \softmax\left(\Z_i^*\right)_j }\right)^2 } \right]  \\
	& \le &  \sum_{l = 1}^{\infty}(n+p)^{-l}  \le \frac2{n+p}.
\end{eqnarray*}

Note that by the conditions that $N > (n+p)\log (n+p)$, these exists some constant $C_2>0$ such that
\textcolor{black}{
	\begin{eqnarray*}
&& E(n,p,r) \\
 &\le & C_2\frac{\beta_{\R} \vee \left(\max_{i,j}\softmax(\Z_i^*)_jp\right) }{\left(\min_{i,j}\softmax(\Z_i^*)_jp\right)^2 }\cdot \frac{ \log (n+p)}{N} \sum_{i=r+1}^{n \wedge p} \sigma_i\left(\Z^*\right)\nonumber  \\
	&  &	+\textcolor{black}{C_2 \frac{\left(\max_{i,j}\softmax(\Z_i^*)_jp \right)^2 \cdot \left[  \beta_{\R} \vee \left(\max_{i,j}\softmax(\Z_i^*)_jp\right) \right]}{ \alpha_{\R} \left(\min_{i,j}\softmax(\Z_i^*)_jp\right)^3 } \cdot \frac{r(n+p) \log (n+p)}{N}  }, 
		\end{eqnarray*}} which completes the proof.
	

\textbf{Proof of Lemma \ref{lem:lem 3}}
By using Taylor expansion, we can rewrite KL divergence as
\begin{eqnarray*}
	&&\sum_{i=1}^n D_{KL} \left(\softmax \left(\Z_i^* \right), \softmax\left(  \hat \Z_i\right)  \right) \\
	& = & \sum_{i=1}^n \left[    \log \sum_{k=1}^pe^{\hat z_{ik}}  - \log \sum_{k=1}^pe^{ z_{ik}^*} - \sum_{j=1}^p\softmax \left(\Z_i^* \right)_j \left(\hat z_{ij} - z_{ij}^* \right) \right] \\
	& = & - \sum_{i=1}^n \left(\hat \Z_{i} - \Z_{i}^* \right)^T \nabla^2 \softmax\left( (1-t) \Z_i^*+t\hat \Z_i\right)\left(\hat \Z_{i} - \Z_{i}^* \right),
\end{eqnarray*}
where $ \nabla^2 \softmax\left( \xi\right) =  - \diag\left\{\softmax(\xi) \right\} + \softmax(\xi)^T\softmax(\xi)  $. Let us denote $\u \triangleq \softmax\left( \xi\right)$ and then for any $\x \in \Reals^{p}\setminus\{\0_p\}$ such that $\x^T\1_p =0$,

\begin{eqnarray*}
	&& \frac{-\x^T \nabla^2 \softmax\left( \xi\right)\x}{\x^T\x} = \frac{\x^T\left(\diag(\u) -\u\u^T \right)\x}{\x^T\x} \\
	&=&  \frac{\left(\diag(\u)^{\frac12}\x\right)^T\left(\I_p -\diag(\u)^{-\frac12}\u\left[\diag(\u)^{-\frac12}\u\right]^T \right)\diag(\u)^{\frac12}\x}{\x^T\x}  \\
	& = & \frac{\left(\diag(\u)^{\frac12}\x\right)^T\left(\I_p -\diag(\u)^{-\frac12}\u\left[\diag(\u)^{-\frac12}\u\right]^T \right)\diag(\u)^{\frac12}\x}{\left(\diag(\u)^{\frac12}\x\right)^T\diag(\u)^{\frac12}\x} \cdot \\
	&& \frac{\left(\diag(\u)^{\frac12}\x\right)^T\diag(\u)^{\frac12}\x}{\x^T\x} 
\end{eqnarray*}
and thus
\begin{eqnarray*}
	&& \min_j  U_J \left( \xi\right)_j\\ & = & \min_j u_j  \cdot \inf_{\y^T\1_p=0}\frac{\y^T \left(\I_p -\diag(\u)^{-\frac12}\u\left[\diag(\u)^{-\frac12}\u\right]^T \right)\y}{\y^T\y}\\
	& \le &\frac{-\x^T \nabla^2 \softmax\left( \xi\right)\x}{\x^T\x} \\
	& \le & \max_ju_j \cdot \sup_{\y^T\1_p=0}\frac{\y^T \left(\I_p -\diag(\u)^{-\frac12}\u\left[\diag(\u)^{-\frac12}\u\right]^T \right)\y}{\y^T\y} \\
	& = & \max_j  \softmax\left( \xi\right)_j.
\end{eqnarray*}

As a result, we have \eqref{eq:equivalence distance}.


\textbf{Proof of Lemma \ref{lem:lem 4}}
According to \eqref{eq:nabla of MLE rewrite}, we rewrite $\displaystyle\nabla \cL_{N}\left( \Z^*\right) = -\frac1N\sum_{k=1}^N\Y^{(k)}$   with $\Y^{(k)} \triangleq  \begin{bmatrix} \frac{\e_{1}(n)^T\E_k \cdot\nabla {clr}^{-1}(\Z_1^*) }{\langle \softmax(\Z_1^*)^T,\E_k^T\e_{1}(n) \rangle} \\ \vdots \\ \frac{\e_{n}(n)^T\E_k \cdot\nabla {clr}^{-1}(\Z_n^*) }{\langle \softmax(\Z_n^*)^T, \E_k^T\e_{n}(n) \rangle} 
\end{bmatrix} $. Notice
\begin{lem}
	 $\E\Y^{(k)}= \0_{n \times p}$.
\end{lem}
\begin{proof}{\allowdisplaybreaks\begin{eqnarray*}
		\expc	\left(\Y^{(k)}_{i}\right) & = &  \expc \left[ \frac{\e_{i}(n)^T \E_k \cdot\nabla {clr}^{-1}(\Z_i^*) }{\langle \softmax(\Z_i^*)^T, \E_k^T\e_{i}(n)\rangle}\right] \\
		& = & \expc \left[\left.\frac{\e_{i}(n)^T \E_k \cdot\nabla {clr}^{-1}(\Z_i^*) }{\langle \softmax(\Z_i^*)^T, \E_k^T\e_{i}(n)\rangle} \right| \e_{i}(n)^T\E_k\1_p =1  \right]\cdot \Prob(\e_{i}(n)^T\E_k\1_p = 1) ,
\end{eqnarray*}}where $\expc \left[\left.\frac{\e_{i}(n)^T \E_k \cdot\nabla {clr}^{-1}(\Z_i^*) }{\langle \softmax(\Z_i^*)^T, \E_k^T\e_{i}(n)\rangle} \right| \e_{i}(n)^T\E_k\1_p =1  \right]$ is just having $\xi \triangleq \e_{i}(n)^T\E_k$ as multinomial distribution with $\softmax\left(\Z_i^*\right)$ as true composition and $1$ as total count; that is,
\begin{eqnarray*}
	&&  \expc \left[\left.\frac{\e_{i}(n)^T \E_k \cdot\nabla {clr}^{-1}(\Z_i^*) }{\langle \softmax(\Z_i^*)^T, \E_k^T\e_{i}(n)\rangle} \right| \e_{i}(n)^T\E_k\1_p =1  \right] \\
	& = & \expc_{\xi \in \{0,1\}^{1\times p} \sim Mult\left(1, \softmax\left(\Z_i^*\right)\right)} \left[\frac{\xi \cdot\nabla {clr}^{-1}(\Z_i^*) }{\langle \softmax(\Z_i^*),\xi \rangle}  \right] \\
	& = & \sum_{j=1}^p  \softmax(\Z_i^*)_j\cdot \frac{\e_{j}(p)^T\nabla {clr}^{-1}(\Z_i^*)}{ \softmax(\Z_i^*)_j} \\
	&= &\sum_{j=1}^p  \e_{j}(p)^T\nabla {clr}^{-1}(\Z_i^*) = \1_p^T\nabla {clr}^{-1}(\Z_i^*) =   \0_p^T.
\end{eqnarray*}

As a result, $\E\Y^{(k)}= \0_{n \times p}$.
\end{proof}
Next we are to use Lemma 6 in \cite{cao2020multisample}, for which we have to provide upper bounds for $\left\| \Y^{(k)}\right\|_2$, $\left\| \expc \left[\Y^{(k)}\right]^T\Y^{(k)}  \right\|_2$, $\left\| \expc  \Y^{(k)} \left[\Y^{(k)}\right]^T\right\|_2$:
\begin{enumerate}
	\item As for $\left\| \Y^{(k)}\right\|_2$, 
	\begin{eqnarray}
	\left\| \Y^{(k)}\right\|_2^2 & \le & \max_{ij} \left\| \frac{\e_{j}(p)^T \cdot\nabla {clr}^{-1}(\Z_i^*) }{\langle \softmax(\Z_i^*)^T,\e_{j}(p)\rangle}\right\|_2^2 = \max_{ij} \left\|\e_{j}(p)  -  {clr}^{-1}(\Z_i^*)^T \right\|_2^2  \nonumber \\
	& = & 1 + \max_{i\in [n]}\left[  \left\|{clr}^{-1}(\Z_i^*)\right\|_2^2 - 2 \min_{j \in [p]}{clr}^{-1}(\Z_i^*)_j \right]. \label{eq:upper bound l2 norm Yk}
	\end{eqnarray}
	\item Speaking of $\left\| \expc \left[\Y^{(k)}\right]^T\Y^{(k)} \right\|_2$,
	{\allowdisplaybreaks	\begin{eqnarray*}
			&&	\expc \left[\Y^{(k)}\right]^T\Y^{(k)} = \begin{bmatrix} \frac{\e_{1}(n)^T\E_k \cdot\nabla {clr}^{-1}(\Z_1^*) }{\langle \softmax(\Z_1^*),\e_{n}(n)^T \E_k\rangle} \\ \vdots \\ \frac{\e_{n}(n)^T\E_k \cdot\nabla {clr}^{-1}(\Z_n^*) }{\langle \softmax(\Z_n^*)^T, \E_k^T\e_{n}(n) \rangle} 
			\end{bmatrix}^T \begin{bmatrix}\expc \frac{\e_{1}(n)^T\E_k \cdot\nabla {clr}^{-1}(\Z_1^*) }{\langle \softmax(\Z_1^*)^T,\E_k^T\e_{1}(n) \rangle} \\ \vdots \\ \frac{\e_{n}(n)^T\E_k \cdot\nabla {clr}^{-1}(\Z_n^*) }{\langle \softmax(\Z_n^*)^T, \E_k^T\e_{n}(n) \rangle} 
			\end{bmatrix} \\
			& = & \sum_{i=1}^n \expc\sum_{s=1}^n \left[\frac{\e_{i}(n)^T \E_k \cdot\nabla {clr}^{-1}(\Z_i^*) }{\langle \softmax(\Z_i^*)^T, \E_k^T\e_{i}(n)\rangle}\right]^T \frac{\e_s^T \E_k \cdot\nabla {clr}^{-1}(\Z_s^*) }{\langle \softmax(\Z_s^*),\e_s^T \E_k\rangle} \\
			& = &  \sum_{i=1}^n \expc\left\{\left[\frac{\e_{i}(n)^T \E_k \cdot\nabla {clr}^{-1}(\Z_i^*) }{\langle \softmax(\Z_i^*)^T, \E_k^T\e_{i}(n)\rangle}\right]^T\left[\frac{\e_{i}(n)^T \E_k \cdot\nabla {clr}^{-1}(\Z_i^*) }{\langle \softmax(\Z_i^*)^T, \E_k^T\e_{i}(n)\rangle}\right]\right\} \\
			& = & \sum_{i=1}^n\Prob\left( \e_{i}(n)^T \E_k \1_p = 1\right)  \cdot  \\
			&& \expc_{\xi \in \{0,1\}^{1\times p} \sim Mult\left(1, \softmax\left(\Z_i^*\right)\right)}\left\{\frac{ \nabla {clr}^{-1}(\Z_i^*)\xi^T\xi\nabla {clr}^{-1}(\Z_i^*)}{\langle \softmax(\Z_i^*),\xi\rangle^2 } \right\} \\
			&= &  \sum_{i=1}^n\nu_i \sum_{j =1 }^p \softmax\left(\Z_i^*\right)_j \left(\e_j(p) - \softmax\left(\Z_i^*\right)^T\right)\left(\e_j(p)^T - \softmax\left(\Z_i^*\right) \right) \\
			&= & \sum_{i=1}^n\nu_i \left\{ \left( \X_{i}^*\right)^T \X_{i}^* +\diag\{X_{ij}^* \}_{j\in[p]}-  \sum_{j=1}^p \left[    \e_j(p) \X_{i}^*+\left( \e_j(p)X_{i}^*\right)^T\right] \right\},	\end{eqnarray*}
	}
	hence 
	\begin{equation}
	\left\| \expc \left[\Y^{(k)}\right]^T\Y^{(k)}  \right\|_2 \le 1- \sum_{i=1}^n\nu_i \left\| \softmax\left(\Z_i^* \right)  \right\|_2^2. \label{eq:upper bound l2 norm of E YTY}
	\end{equation}
	\item Lastly, for $\left\| \expc\Y^{(k)} \left[\Y^{(k)}\right]^T \right\|_2$, 
	{\allowdisplaybreaks	\begin{eqnarray*}
			&& \expc\Y^{(k)} \left[\Y^{(k)}\right]^T=
			\diag \left\{ \expc \frac{\e_{i}(n)^T \E_k \cdot\left[\nabla {clr}^{-1}(\Z_i^*)\right]^2  \E_k^T\e_{i}(n)}{\langle \softmax(\Z_i^*)^T, \E_k^T\e_{i}(n)\rangle^2 } \right\} \\
			& = & \diag \left\{ \Prob\left( \e_{i}(n)^T \E_k \1_p = 1\right)  \cdot \right. \\
			&& \left. \expc \left[\left. \frac{\e_{i}(n)^T \E_k \cdot\left[\nabla {clr}^{-1}(\Z_i^*)\right]^2  \E_k^T\e_{i}(n)}{\langle \softmax(\Z_i^*)^T, \E_k^T\e_{i}(n)\rangle^2 } \right|\e_{i}(n)^T \E_k \1_p = 1\right] \right. \\
			& = & \diag \left\{\nu_i \expc_{\xi \in \{0,1\}^{1\times p} \sim Mult\left(1, \softmax\left(\Z_i^*\right)\right)} \left[\frac{\xi \cdot\left[\nabla {clr}^{-1}(\Z_i^*)\right]^2  \xi^T}{\langle \softmax(\Z_i^*),\xi\rangle^2 } \right] \right\} \\
			& = & \diag \left\{ \nu_i \cdot \sum_{j=1}^p \softmax(\Z_i^*)_j \cdot  \frac{\e_{j}(p)^T \cdot\left[\nabla {clr}^{-1}(\Z_i^*)\right]^2  \e_{j}(p)}{\left[ \softmax(\Z_i^*)_j\right]^2 } \right\}\\
			& = & \diag \left\{ \nu_i \sum_{j=1}^p \softmax(\Z_i^*)_j \cdot \left\|\e_j(p) - \softmax(\Z_i^*)^T\right\|^2 \right\} \\
			& = & \diag \left\{ \nu_i \left( 1 -  \left\| \softmax(\Z_i^*)\right\|_2^2  \right) \right\},
	\end{eqnarray*}}
	Consequentially,
	\begin{equation}
	\left\| \expc\Y^{(k)} \left[\Y^{(k)}\right]^T \right\|_2=  \max_{i \in [p]}  \nu_i \left( 1 -  \left\| \softmax(\Z_i^*)\right\|_2^2  \right) . \label{eq:upper bound l2 norm of EYYT}
	\end{equation}
\end{enumerate}
By applying Lemma 6 in \cite{cao2020multisample}, \eqref{eq:upper bound l2 norm Yk}, \eqref{eq:upper bound l2 norm of E YTY}, \eqref{eq:upper bound l2 norm of EYYT} imply
\begin{eqnarray}
\Prob\left( \left\| \frac1N\sum_{k=1}^N Y_k\right\|_2 \ge t \right) \le (n+p) \exp \left( - \frac{N^2t^2/2}{ \sigma^2 + MNt/3} \right),
\end{eqnarray}
where $ M,\sigma^2 $ are in Lemma \ref{lem:lem 4}. Since Lemma 6 in \cite{cao2020multisample} implies
\begin{equation} \left\| \frac1N\sum_{k=1}^N Y_k\right\|_2  \le \left[\frac{M}3 + \sqrt{\frac{M^2}9 + \frac{\sigma^2}{ \log (n+p)}}\right]\frac{2 \log (n+p)}N,
\end{equation}
with probabity at least $1 - \frac1{n+p}$.

\subsection{Concentration inequalities}
\begin{lem}
	\label{lem:lem 7}
	Let $n \times p$ random matrices $\{\E_k\}_{k=1}^N$ be independent and identically distributed with distribution $\Pi = \R\1_p^T \circ \softmax(\Z^*)$ on $\{\e_i(n) \e_j(p)^T, (i,j) \in [n]\times [p] \}$ and $\{\epsilon_k\}_{k=1}^N$ is an i.i.d. Rademacher sequence. Assume $\frac{\alpha_{\R}}{n} \le R_i \le \frac{\beta_{\R}}{n}$ for any $\Z^* \in \cS$ we have the upper bound 
\begin{eqnarray*}
&& \expc \left\| N^{-1} \sum_{k=1}^N \epsilon_k  \E_k \right\|_2  \\
&\le &\left\{\frac{28 \log (n+p) \left[(\beta_{\R}/n)\vee  \max_{i,j}\softmax(\Z_i^*)_j \right]}N  \right\}^{\frac12 }  + \frac{28\log(n+p)}N.
\end{eqnarray*}

\end{lem}
\begin{lem}
	\label{lem:lem 8}
	We define a constraint set $\cD (T) $ with some constant $T>0$, 
	\begin{equation}
	\cD (T) \triangleq \left\{\Z \left| \sum_{i=1}^n D_{KL}\left(\softmax\left(\Z_i^*\right),\softmax(\Z_i)\right)  \le T \right.\right\}.
	\end{equation}
	And denote by $Z_T$ the function on the constraint set $\cD(T)$
	$$
	U_T \triangleq \sup_{\Z \in \cD(T)}\left| \cL_N \left(\hat \Z\right) - \cL_N \left( \Z^* \right) - \sum_{i=1}^n R_i D_{KL}\left(\softmax\left(\Z_i^*\right),\softmax(\hat \Z_i)\right)   \right|.
	$$
\end{lem}

Under the assumption that $\frac{\alpha_{\R}}{n} \le R_i \le \frac{\beta_{\R}}{n}$, if $\hat \Z $ satisfies 
$
\left\| \hat \Z - \Z^* \right\|_* \le 4 \sqrt{2r} \left\| \hat \Z - \Z^* \right\|_F + 4 \sum_{i=r+1}^{n \wedge p} \sigma_i\left(\Z^*\right)$, then 
\begin{equation}
\Prob \left( U_T \ge \frac{\alpha_{\R}T}{n} +E(n,p,r) \right) \le \exp\left[ - \frac{\alpha_{\R}^2 NT^2 }{512 \left(n \log \frac{\max_{i,j} \softmax\left(\Z_i^* \right)_j }{\min_{i,j} \softmax\left(\Z_i^*\right)_j }\right)^2 } \right],
\label{eq:lem8}
\end{equation}
where $E(n,p,r)$ is defined in \eqref{eq:E(n,p,r)}.
\begin{proof}
	Since 
	$$\sup_{\hat \Z, \Z^* \in \cS } \max_{i,j}\left| \log \softmax\left(\Z_i^* \right)_j - \log \softmax\left( \hat  \Z_i \right)_j \right| \le \log \frac{\max_{i,j} \softmax\left(\Z_i^* \right)_j }{\min_{i,j} \softmax\left(\Z_i^*\right)_j },$$
	we obtain the following concentration inequality by a version ho Hoeffding's inequality due to Theorem 14.2 of \cite{buhlmann2011statistics},
\begin{equation}\label{eq:concentration inequality UT}
\Prob\left(U_T - \expc U_T \ge \frac{\alpha_{\R} T}{8n} \right) \le \exp \left[- \frac{\alpha_{\R}^2 NT^2}{512 \left(n \log \frac{\max_{i,j} \softmax\left(\Z_i^* \right)_j }{\min_{i,j} \softmax\left(\Z_i^*\right)_j }\right)^2 } \right].
\end{equation}

It remains to upper bound the quantity $\expc U_T$. By using a standard symmetrization argument, we obtain
\begin{eqnarray*}
	\expc U_t &= &\expc \sup_{\Z \in \cD(T)}\left| \cL_N \left(\hat \Z\right) - \cL_N \left( \Z^* \right) - \sum_{i=1}^n R_i D_{KL}\left(\softmax\left(\Z_i^*\right),\softmax(\hat \Z_i)\right)   \right| \\
	& \le & 2 \expc \left( \sup_{\Z \in \cD(T)} \left| \cL_N \left(\hat \Z\right) - \cL_N \left( \Z^* \right) \right| \right) \\ 
	&=& 2 \expc \left( \sup_{\Z \in \cD(T)} \left| N^{-1} \sum_{k=1}^N \epsilon_k \left\langle \log \softmax\left(\Z^* \right) - \log \softmax\left(\hat \Z \right) , \E_k\right\rangle\right| \right) \\
	& = & 2 \expc \left( \sup_{\Z \in \cD(T)} \left| N^{-1} \sum_{k=1}^N \epsilon_k \sum_{i,j }\1_{\{\E_k = \e_i(n)\e_j(p)^T\}} \log \frac{\softmax\left(\Z_i^* \right)_j}{\softmax\left(\hat \Z_i \right)_j} \right| \right),
\end{eqnarray*}
where $\{\epsilon_k\}_{k=1}^N$ is an independent and identically distributed Rademacher sequence. Then the contraction principle from Theorem 4.12 in \cite{ledoux2013probability}, together with Holder's inequality between nuclear and operator norm, yields
\begin{eqnarray*}
\expc U_T& \le &  \frac{4}{\min_{i,j} \softmax\left(\Z_i^* \right)_j} \\
 & & \cdot \expc \left( \sup_{\Z \in \cD(T)} \left| N^{-1} \sum_{k=1}^N \epsilon_k \left\langle  \log \softmax\left(\Z^* \right) - \log \softmax\left(\hat \Z \right) , \E_k \right\rangle \right| \right) \\
 & \le & \frac{4}{\min_{i,j} \softmax\left(\Z_i^* \right)_j}  \sup_{\Z \in \cD(T)} \left\| \Z^* - \hat \Z\right\|_* \cdot \expc \left\| N^{-1} \sum_{k=1}^N \epsilon_k  \E_k \right\|_2.
\end{eqnarray*}

We bound $\displaystyle \expc \left\| N^{-1} \sum_{k=1}^N \epsilon_k  \E_k \right\|_2$ by applying Lemma \ref{lem:lem 7}. Under the assumption that $
\left\| \hat \Z - \Z^* \right\|_* \le 4 \sqrt{2r} \left\| \hat \Z - \Z^* \right\|_F + 4 \sum_{i=r+1}^{n \wedge p} \sigma_i\left(\Z^*\right)$, applying Lemma \ref{lem:lem 3}, we can bound $\left\| \hat \Z - \Z^* \right\|_*$ by 
\begin{eqnarray*}
&& \sup_{\hat \Z \in \cD(T)}\left\| \hat \Z - \Z^* \right\|_* \\
& \le & 4 \sum_{i=r+1}^{n \wedge p} \sigma_i\left(\Z^*\right) +  4 \sqrt{2r} \sup_{\hat  \Z \in \cD(T)} \left\| \hat \Z - \Z^* \right\|_F \\
& \le & 4 \sum_{i=r+1}^{n \wedge p} \sigma_i\left(\Z^*\right) +  \frac{8 \max_{i,j}\softmax \left( \Z_i^*\right)_j \sqrt{r}}{\sqrt{\min_{i,j}\softmax \left( \Z_i^*\right)_j }} \sqrt{\sum_{i=1}^n D_{KL}\left(\softmax\left(\hat{\Z}_i\right),\softmax(\Z_i^*)\right) } \\
& \le & 4 \sum_{i=r+1}^{n \wedge p} \sigma_i\left(\Z^*\right) +  \frac{8 \max_{i,j}\softmax \left( \Z_i^*\right)_j \sqrt{rT}}{\sqrt{\min_{i,j}\softmax \left( \Z_i^*\right)_j }}.
\end{eqnarray*}

As a result,
\begin{eqnarray*}
\expc U_T  &\le &   \frac{16}{ \min_{i,j}\softmax \left( \Z_i^*\right)_j  }\\
&& \cdot \left[\left\{\frac{28 \log (n+p) \left[(\beta_{\R}/n)\vee  \max_{i,j}\softmax(\Z_i^*)_j \right]}N  \right\}^{\frac12 }  + \frac{28\log(n+p)}N\right] \\
&& \cdot \left[ \sum_{i=r+1}^{n \wedge p} \sigma_i\left(\Z^*\right) +  \frac{2 \max_{i,j}\softmax \left( \Z_i^*\right)_j \sqrt{rT}}{\sqrt{\min_{i,j}\softmax \left( \Z_i^*\right)_j }}\right] \\
&\le &  E(n,p,r) + \frac{\alpha_\R T}{8n}.
\end{eqnarray*}

finally, plugging the upper bound of $\expc U_t$ into concentration inequality \eqref{eq:concentration inequality UT}, we obtain \eqref{eq:lem8}.
\end{proof}

\textbf{Proof of main results}
%

\subsection{Proof of upper bound on singular subspace distance
}

\textbf{A simple proof of the upper bound}

\begin{lem}[Weyl's lemma] \label{eq:Weyl's lemma}

	$\left|  \sigma_{i} \left(\hat \Z \right)- \sigma_{i} \left( \Z^* \right) \right| \le \left\| \hat \Z - \Z^* \right\| $.
\end{lem}
Weyl's lemma \ref{eq:Weyl's lemma} and Davis-Kahan $\sin \Theta$ theorem (see a version from \cite{li2018two,hsu2016COMS4772}), for right singular vectors $\V_{\Z^*}, \V_{\hat \Z}$ and an unitarily invariant norm $\|\cdot \|_{\cU }$ we obtain 
$$
\left\| \sin \Theta \left( \V_{\hat \Z},  \V_{ \Z^*}\right) \right\|_{\cU} \le \frac{\left\| \left( \hat \Z -  \Z^* \right) \V_{\Z^*} \right\|_{\cU}}{\sigma_r\left(\Z^* \right) - \sigma_{r+1}\left(\hat \Z \right)} 
 \le \text{ \eqref{eq:lower bound of r th largest singular value} }  \frac{2\left\| \left( \hat \Z -  \Z^* \right) \V_{\Z^*} \right\|_{\cU}}{\sigma_r\left(\Z^* \right) },
$$
and if we pick $\|\cdot \|_{\cU } = \|\cdot \|_{F }$,
Theorem \ref{thm:thm1} implies Theorem \ref{thm:principal subspace estimation}. Same for left singular vectors of course.


\textbf{Proofs of the asymptotic expansion of the singular subspace distance}
\label{append:proofs of asymptotic expansions of the singular subspace distances}

As an extension, we can also derive Theorem \ref{thm:principal subspace estimation} from following result similar to asymptotic expansion results under Frobenius norm in \cite{koltchinskii2014concentration,koltchinskii2017normal,li2018two,xia2018confidence}:

\begin{thm}
	\label{thm:principal subspace estimation asymptotic}
	
	Under assumptions and selection of tuning parameter in Theorem \ref{thm:thm1} and by furtherly imposing a lower bound on $r$th largest singular value \eqref{eq:lower bound of r th largest singular value}, then we have
	\begin{eqnarray*}
		&&	\frac{ \sigma_r\left( \Z \right)}{ \left\| \hat \Z - \Z^* \right\|_{\cU } }	\left\| \sin \Theta \left( \V_{\hat \Z},  \V_{ \Z^*}\right) \right\|_{F}\\
		& = & 	\frac{ \sigma_r\left( \Z \right)}{ \left\| \hat \Z - \Z^* \right\|_{\cU } } \left\| \left( \hat \Z -  \Z^*\right)  \V_{ \Z^*} \diag\left\{ \sigma_1 \left( \Z^*\right) ,\ldots, \sigma_r \left( \Z^* \right)\right\}^{-1}\right\|_{F} \\
		&&+ O_{P}\left(r^{-\frac12}\frac{ \left\| \hat \Z - \Z^* \right\|_{\cU } }{ \sigma_r\left( \Z^* \right)}  \log (n)\right).
	\end{eqnarray*}
\end{thm}

	\end{document}